\documentclass{article}
\usepackage[utf8]{inputenc}
\usepackage[T1]{fontenc}
\usepackage{geometry}
\usepackage{amsmath}
\usepackage{amsfonts}
\usepackage{array}
\usepackage{babel}
\usepackage{bbm}
\usepackage{graphicx}
\usepackage{subcaption}
\usepackage{float}
\usepackage{booktabs}
\usepackage{multirow} 
\usepackage{url}
\usepackage{multicol}
\usepackage[square,numbers]{natbib}
\usepackage{color}
\usepackage[dvipsnames]{xcolor}
\usepackage{setspace}

\usepackage{authblk}

\usepackage{algorithm}
\usepackage[noend]{algpseudocode}

\definecolor{emerald}{RGB}{12, 166, 151}

\newcommand{\bX}{\mathbf{X}}

\usepackage{longtable}

\usepackage{afterpage}

\title{Bayesian Safety Surveillance with Adaptive Bias Correction}
\author[1]{Fan Bu}
\author[1,2]{Martijn J. Schuemie}
\author[3]{Akihiko Nishimura}
\author[4,5]{Louisa H. Smith}
\author[5]{Kristin Kostka}
\author[6]{Thomas Falconer}
\author[6]{Jody-Ann McLeggon}
\author[2]{Patrick B. Ryan}
\author[6]{George Hripcsak}
\author[1]{Marc A. Suchard}

\affil[1]{Department of Biostatistics, University of California, Los Angeles}
\affil[2]{Janssen Research and Development}
\affil[3]{Department of Biostatistics, Bloomberg School of Public Health, Johns Hopkins University}
\affil[4]{Department of Health Sciences, Northeastern University}
\affil[5]{The OHDSI Center at the Roux Institute, Northeastern University}
\affil[6]{Department of Biomedical Informatics, Columbia University}

\date{}

\begin{document}

\maketitle

\begin{abstract}
    Post-market safety surveillance is an integral part of mass vaccination programs. 
    Typically relying on sequential analysis of real-world health data as they accrue, safety surveillance is challenged by the difficulty of sequential multiple testing and by biases induced by residual confounding in observational health data. 
    The current standard approach based on the maximized sequential probability ratio test (MaxSPRT) fails to satisfactorily address these practical challenges and it remains a rigid framework that requires pre-specification of the surveillance schedule. 
    We develop an alternative Bayesian surveillance procedure that addresses both the aforementioned challenges of sequential testing and residual biases using a more flexible framework. 
    We adopt a joint statistical modeling approach to sequentially estimate the effect of vaccine exposure on the adverse event of interest and correct for estimation bias by simultaneously analyzing a large set of negative control outcomes through a Bayesian hierarchical model. 
    We then compute a posterior probability of the alternative hypothesis via Markov chain Monte Carlo sampling and use it for sequential detection of safety signals. 
    Through an empirical evaluation using six US observational healthcare databases covering more than 360 million patients, 
    we benchmark the proposed procedure against MaxSPRT on testing errors and estimation accuracy, 
    under two epidemiological designs, the historical comparator and the self-controlled case series. 
    We demonstrate that our procedure substantially reduces Type 1 error rates, maintains high statistical power and fast signal detection, and provides considerably more accurate estimation. 
    Given the extensiveness of the empirical study which yields more than 7 million sets of results, we present all results in a public \texttt{R} ShinyApp. 
    As an effort to promote open science, we provide full implementation of our method in the open-source \texttt{R} package \texttt{EvidenceSynthesis}. 
\end{abstract}

\section{Introduction}

Mass vaccination is a cornerstone of effective disease control \citep{anderson1982directly,francis1982prevention,centers2005epidemiology,nowak2015promoting}. 
Addressing any safety concerns over new vaccine products is therefore essential. 
Despite pre-approval clinical trials that ensure vaccine effectiveness and safety, rare but severe adverse events often go undetected due to limited sample sizes and statistical power. 
Post-approval safety surveillance is thus a critical component of mass vaccination programs. 
One common approach of safety surveillance is to sequentially analyze real-world observational data as they accrue over time
\citep{verstraeten2003vaccine,lieu2007real,
baker2013post, leite2016near,
moro2016surveillance,lee2020postapproval,EMA-covid-plan, vaer2021, vsd2021,cisa2021,who2021covid}.

However, there are two major challenges in safety surveillance using observational data. 
One challenge is to adjust for sequential multiple testing with a flexible surveillance schedule while allowing for rapid detection of true safety signals. 
Another challenge is to account for and correct for residual systematic error in the observational data that induces bias and can inflate decision errors \citep{rodrigues1999use,glanz2006four,newcomer2018bias}. 
The maximized sequential probability ratio test (MaxSPRT) stands as a standard and commonly adopted statistical framework for sequential safety surveillance \citep{kulldorf2011}.  
MaxSPRT, however, does not provide a fully satisfactory solution to these challenges \citep{schuemie2022vaccine,schuemie2023adjusting}. 
MaxSPRT remains a rigid framework that requires a pre-fixed surveillance schedule and does not allow the analysis plan to adapt to real-world data accrual behaviors. 
Morevoer, MaxSPRT does not naturally allow for extensions to correct for biases in observational studies. 

In this work, we propose an alternative safety surveillance procedure that addresses these two challenges in a unified and interpretable framework. 
We develop the procedure based on Bayesian sequential analysis, with an empirical modeling component that adaptively corrects for biases. 
More specifically, at each sequential analysis timepoint, we perform Bayesian inference to obtain a posterior distribution for the effect of vaccination on the adverse event of interest; this is augmented by an adaptive bias correction approach that builds on an empirical bias distribution learned through analysis of negative control outcomes. 
From the posterior distribution, we compute the posterior probability of the hypothesis that vaccination increases the risk of the adverse event, given the data accrued. 
We then use the posterior probability as a test statistic for signal detection: at the first timepoint when the posterior probability exceeds a user-specified threshold, we declare a safety signal.  
Such surveillance procedure, unlike MaxSPRT, does not require a pre-specified surveillance schedule, but instead only relies on data evidence that has already accrued. 
We will detail our proposed framework in Section~\ref{sec: methods}.

Through comprehensive empirical evaluations using large-scale observational healthcare databases of more than 360 million unique patients, we demonstrate the benefits of our methods. 
Compared to the standard approach, our Bayesian surveillance procedure offers greater flexibility in the surveillance schedule, improved transparency and interpretability for evidence extraction and decision-making, and more reliable error control through bias correction.  
In the remainder of this section, we provide necessary background information about safety surveillance as a statistical problem. 

\subsection{The statistical problem: sequential hypothesis testing}
\label{sec: sequential-test-background}
In safety surveillance, the key question we wish to answer is: for a particular vaccine (e.g., seasonal flu vaccine) and an adverse event outcome (e.g. the Guillain-Barré Syndrome), is the incidence rate of this adverse event increased after taking the vaccine compared to the rate in the unvaccinated? 

The key quantity of interest to estimate in order to answer such question is the incidence rate ratio, $RR$, defined as the ratio between the incidence rate during a risk window post vaccination versus that without vaccination.
Thus, safety surveillance is essentially a hypothesis testing problem: 
\begin{equation}
    H_0: RR \leq 1 \qquad \text{ v.s. } \qquad  H_1: RR > 1.
    \label{eq: effect-hypotheses}
\end{equation}
Here, the alternative hypothesis $H_1$ indicates an increased risk of the adverse event due to vaccine exposure and thus implies a safety concern (or a safety signal). 

Oftentimes we perform inference on the natural logarithm of $RR$, i.e., with quantity of interest 
$\beta = \log(RR)$. Then the equivalent hypotheses are:
\begin{equation}
    H_0: \beta \leq 0 \qquad \text{ v.s. } \qquad  H_1: \beta > 0.
    \label{eq: log-effect-hypotheses}
\end{equation}

In the context of safety surveillance, as data accrue over time, we need to conduct the hypothesis test \eqref{eq: effect-hypotheses} or \eqref{eq: log-effect-hypotheses} sequentially over time as well. 
Typically this requires updating a test statistic $W_t$ at each analysis timepoint $t$ and compare the statistic against a decision threshold. 
As soon as $W_t$ exceeds the threshold, we reject $H_0$ in favor of $H_1$ and declare a safety signal. 



\subsection{The data}
\label{sec: data-background}
Safety surveillance usually relies on observational healthcare databases including 
administrative claims, electronic health records (EHRs), local registries and spontaneous reports.
We perform analysis at discrete time points when a new batch of data has accrued.

Let $t = 1, 2, \ldots, T_{\max}$ denote each time point of analysis, where $T_{\max}$ represents the maximum length of surveillance window ($T$ can be infinite if a maximum length is not pre-set). 
Denote by $\bX_{t}$ \emph{all} the data that have accrued up to $t$ since the beginning of the surveillance. 
Then, at time $t$, we base analysis on $\bX_{t}$.\footnote{Here, the data $\bX_t$ include the timing and occurrence of vaccination and adverse events, as well as other subject-level covariates available through time $t$.}
For instance, in the case of monthly data accrual, $\bX_3$ should include \emph{all} data available in the first three months.
With respect to sequential hypothesis testing (as described in Section~\ref{sec: sequential-test-background}), the test statistic $W_t$ shall be computed as a function of data $\bX_t$, $W_t := W_t(\bX_t)$, at each analysis timepoint $t$. 

\subsection{MaxSPRT: current standard approach for safety surveillance}

MaxSPRT
is a sequential testing approach developed by \cite{kulldorf2011} to account for sequential multiplicity due to repeated data analyses in safety surveillance. At each analysis time point $t$, the MaxSPRT test statistic is defined as the maximized log likelihood ratio (LLR) between the alternative hypothesis $H_1$ and null hypothesis $H_0$:

\begin{equation}
\label{eq: maxsprt-statistic}
    W_t(\bX_t) := \log(\max \text{LR}_t) = \log\left(\frac{\max_{H_1: \beta > 0} p(\bX_t \mid \beta)}{\max_{H_0: \beta \leq 0} p(\bX_t \mid \beta)}\right),
\end{equation}
where $p(\bX_t \mid \beta)$ denotes the joint probability density function of data $\bX_t$
(used as the likelihood function in regards to $\beta$) specified by the model of choice. 
In \cite{kulldorf2011}, examples of data models include binomial and Poisson distributions for incidence counts of adverse events. 

To calculate a decision threshold, referred to as the ``critical value'' $cv$, the user must pre-specify the surveillance schedule, including the length of surveillance (total number of analyses) and the incremental sample sizes between every two consecutive analyses. 
Then, given a desired significance level $\alpha$ (usually $\alpha=0.05$), a constant $cv$ is numerically computed for a particular data model. 
This can be thought of as a method of ``alpha-spending'' that accounts for the temporal dependency between sequential observations. 

At the first time $t$ at which the LLR statistic exceeds the critical value, i.e., $\log(\max \text{LR}_t) > cv$, $H_0$ is rejected and a safety signal is declared. 

\subsection{Overview of Bayesian inference and sequential analysis}

In a Bayesian framework, inference is based on posterior updates of beliefs from prior beliefs given observed data, which is naturally sequential as we can update beliefs whenever new data are observed. 

Let $\pi_0(\beta)$ denote the \emph{prior} probability density function for $\beta$. 
Given $p(\bX_t \mid \beta)$(the likelihood function), we can obtain the \emph{posterior} distribution $\pi_t(\beta \mid \bX_t)$ using the Bayes rule:
\begin{equation}
\label{eq: belief-update}
\pi_t(\beta \mid \bX_t) \propto p(\bX_t \mid \beta) \pi_0(\beta).
\end{equation}
Here, the symbol ``$\propto$'' indicates that the left hand side is proportional to the right hand side by a factor of constant quantities. 

One important feature of Bayesian hypothesis testing is that we need to specify our \emph{prior beliefs} on the two hypotheses $H_0$ and $H_1$, which can be implied by the prior density function $\pi_0(\beta)$.
For example, a ``neutral'' choice is to specify equal prior probabilities such that $P(H_0) = P(H_1) = \frac{1}{2}$. 

For Bayesian sequential testing, the ``test statistic'' $W_t := W_t(\bX_t)$ commonly takes two forms. 

The first is the Bayes Factor, which accounts for the ratio of marginal data evidence in support of each hypothesis:
\begin{equation*}
    BF_{10}^{(t)} := \frac{m_1(\bX_t)}{m_0(\bX_t)},
\end{equation*}
where $m_i(\bX_t) = \int_{\beta \in H_i} p_t(\bX_t \mid \beta)\pi_0(\beta) d\beta$, for $i=0,1$. 
Common choices for the decision threshold include $10, 20$ and $30$ \citep{barnard1946sequential,wetherill1961bayesian,berger1994unified,berger1988likelihood,berger1997unified,berger1999simultaneous,jha2009bayesian}.
For example, using the Bayes Factor and taking $20$ as the threshold, at the first time $t$ when $BF_{10}^{(t)} > 20$, we should stop the study and reject $H_0$
in favor of $H_1$. 

The second is the posterior probability of either hypothesis. 
A straightforward test statistic $W_t$ for accepting $H_1$ can be its posterior probability $P_{1,t}$ given the data $\bX_t$ accrued up to time $t$:
\begin{equation}
\label{eq: post-prob-calculation}
    W_t(\bX_t) := P_{1,t} = P_t(H_1: \beta > 0 \mid \bX_t)
    = \int_{H_1: \beta > 0} \pi_t(\beta \mid \bX_t) d\beta.
\end{equation}

The decision threshold for the posterior probability can take multiple values between $0$ and $1$. 
Common choices used by previous work include $0.8$, $0.9$, and $0.95$ \citep{cornfield1966bayesian,thall1995bayesian,smith2006implementation,zhou2008bayesian,berry2010bayesian,li2020bayesian}. 


\medskip
\paragraph{Organization of subsequent sections} The rest of this manuscript is structured as follows. 
In Section~\ref{sec: limitations} we discuss the limitations of the standard approach for safety surveillance and motivate the development of an alternative framework. 
We propose our Bayesian procedure in Section~\ref{sec: methods} and then describe a large-scale empirical evaluation to benchmark its performance against MaxSPRT. 
We summarize results of the empirical evaluation in Section~\ref{sec: results}. 
We then briefly investigate the association between varicella zoster (Shingrix) vaccination and occurrences of the Guillain-Barré syndrome in Section~\ref{sec: case-study} and finally conclude with discussions in Section~\ref{sec: discussion}. 

\section{Limitations of existing approaches}
\label{sec: limitations}
In this section, we illustrate the limitations of the existing safety surveillance framework. Through simple examples via simulations, we wish to show that (1) the standard MaxSPRT framework is inflexible and can lead to inconsistent decision-making, and (2) residual systematic error in data can bias analyses and requires additional correction. 

Suppose we wish to learn the effect of a hypothetical vaccine exposure on some hypothetical adverse event outcome. 
We first assume a null outcome that has no relation to the exposure ($RR=1$) to empirically evaluate Type 1 error rates over sequential analyses of data. 
Then we consider a ``positive'' outcome for which $RR=2$ (i.e., the hypothetical vaccine elevates the incidence rate by 2 fold) to examine the accuracy of estimating $RR$ using two common epidemiological designs. 

\subsection{MaxSPRT is inflexible and can produce inconsistent decisions}

A key shortcoming of the MaxSPRT framework lies in its inflexibility --- it requires users to pre-specify the entire surveillance schedule, including the total length of the study, and the incremental sample sizes between analyses. Such inflexibility is inconvenient and can also produce inconsistent decisions between different schedules. 

We showcase this through a simulation experiment where  MaxSPRT is implemented, assuming the Poisson model as the statistical model for incidence counts of the outcome event. 
Consider a surveillance study with monthly data analyses, where during each month approximately $10$ incidents are to be expected, and data collection or accrual ends after $24$ months. 
To emulate this, we simulate outcome incidence counts sequentially such that: (1) between two consecutive data looks, $10$ events for the outcome are expected during the time interval; and (2) by the end of the data accrual process, $10 \times 24 = 240$ total events are expected.

To implement MaxSPRT, the total length of surveillance must be specified \emph{a priori} (note that we have already specified the group sizes between data looks in terms of expected event counts). Consider three different schedules that differ in the surveillance length pre-planned \emph{before seeing the data}:
\begin{itemize}
    \item[A] \textbf{``Oracle''}: \textbf{24} total data looks are pre-planned; this is in agreement with the actual length of data accrual.
    \item[B] \textbf{``Hacky extension''}: \textbf{12} total data looks are pre-planned; but since data are available beyond the pre-planned analysis endpoint, in practice, if no signal is detected within 12 data looks, 
    one might extend the study beyond the original surveillance schedule.
    \item[C] \textbf{``Early stop''}: \textbf{36} total data looks are pre-planned; but since data  are only available up to $24$ data looks, one has to stop the study ``early''.
\end{itemize}
The three different specifications for the surveillance schedule result in  three \emph{different} decision thresholds used by MaxSPRT, and,  consequently, different decisions may be made, even with the \emph{exact same dataset} generated via simulation.

\begin{figure}
	\centering
	\includegraphics[width = 0.7\textwidth]{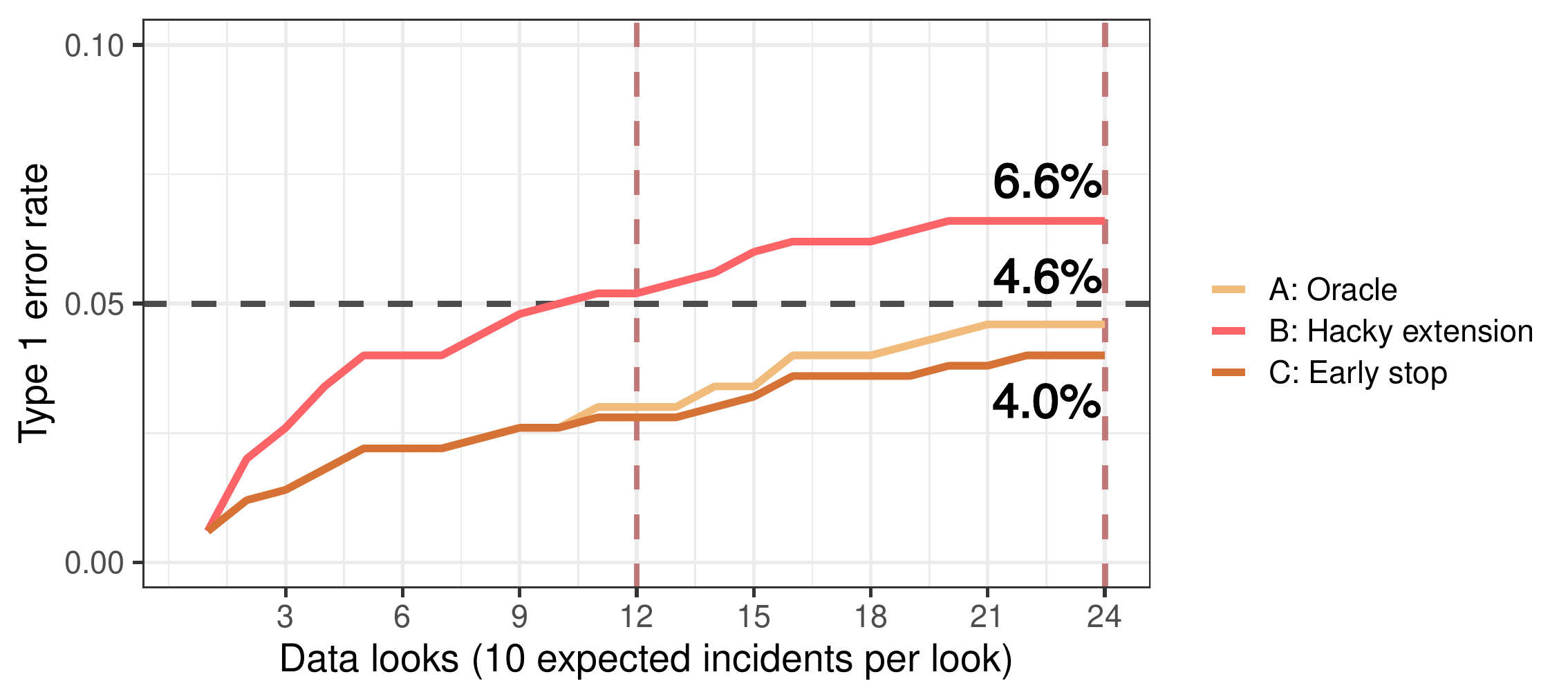}
 \caption{\textbf{Inconsistent decisions by MaxSPRT due to required pre-specified analysis plans.} Empirical Type 1 error rate over 24 monthly surveillance data looks, with \textbf{same} data accrued monthly, only with different analysis schedules. Error rates calculated for 500 simulations in which the synthesized exposure has \textbf{no effect} on the synthesized outcome. 
 The dark gray horizontal line indicates the pre-specified significance level, $\alpha = 5\%$. 
 The end-of-analysis Type 1 error rates (by the end of 24 data looks) are annotated by text.}
 \label{fig: MaxSPRT-3-plans}
\end{figure}

In Figure~\ref{fig: MaxSPRT-3-plans} we plot the empirical Type 1 error rates accumulated over sequential analyses using the three different pre-specified schedules. 
The Type 1 error rate is measured by the fraction of ``reject $H_0$'' decisions across 500 repeated simulations, where MaxSPRT is implemented with a pre-specified significance level $\alpha = 5\%$. 

The differences between the three curves reflect the difficulty of planning ahead in real-world studies. 
Only the ``oracle'' that predicts data collection length correctly and pre-plans surveillance accordingly is able to guarantee a close-to-nominal level of Type 1 error rate.
The ``early stop'' plan effectively does not exhaust the amount of ``$\alpha$'' that is pre-planned to spend over 36 data looks, and thus at 24 total data looks it would be under-powered.
The ``hacky extension'' plan, on the other hand, overshoots on Type 1 error because it has pre-planned for 12 analyses only (note that its Type 1 error rate is controlled at around 5\% by 12 data looks). 
However, under the MaxSPRT framework (or a similar alpha-spending approach), 
if data behavior deviates from the \textit{a priori} prediction (which happens often), 
there is no feasible option to adaptively adjust the analysis plan that allows us to use more available data or preserve power.
Therefore, it is critical to develop a flexible framework that does \emph{not} depend on pre-specification of the surveillance schedule, which we will discuss in Section~\ref{sec: our-method}.



\subsection{Residual systematic error can bias analyses}

We next present a toy simulated example where residual systematic error leads to biased estimation of $RR$. 
Residual systematic error can be thought of as patterns, trends and covariates in the observational data that are not observed or measured, or not adjusted for in the study design or statistical model, but remain common because the data were collected for other uses than the study at hand.

We consider a safety surveillance study that lasts one year with data analysis scheduled at the end of each month (thus 12 total data looks); one year of historical data collected prior to the surveillance study are also available. 
Suppose the hypothetical exposure has a positive effect on a hypothetical outcome with $RR=2$. We use two very common epidemiological designs for vaccine safety surveillance: the historical comparator \citep{belongia2010real,li2021characterising,klein2021surveillance} and the self-controlled case series (SCCS, \cite{Whitaker2006,farrington1995new,farrington2004control}), to estimate $RR$ through sequential analysis and examine the estimation accuracy. 

\begin{figure}
    \centering
    \begin{subfigure}{0.48\textwidth}
        \includegraphics[width=\textwidth]{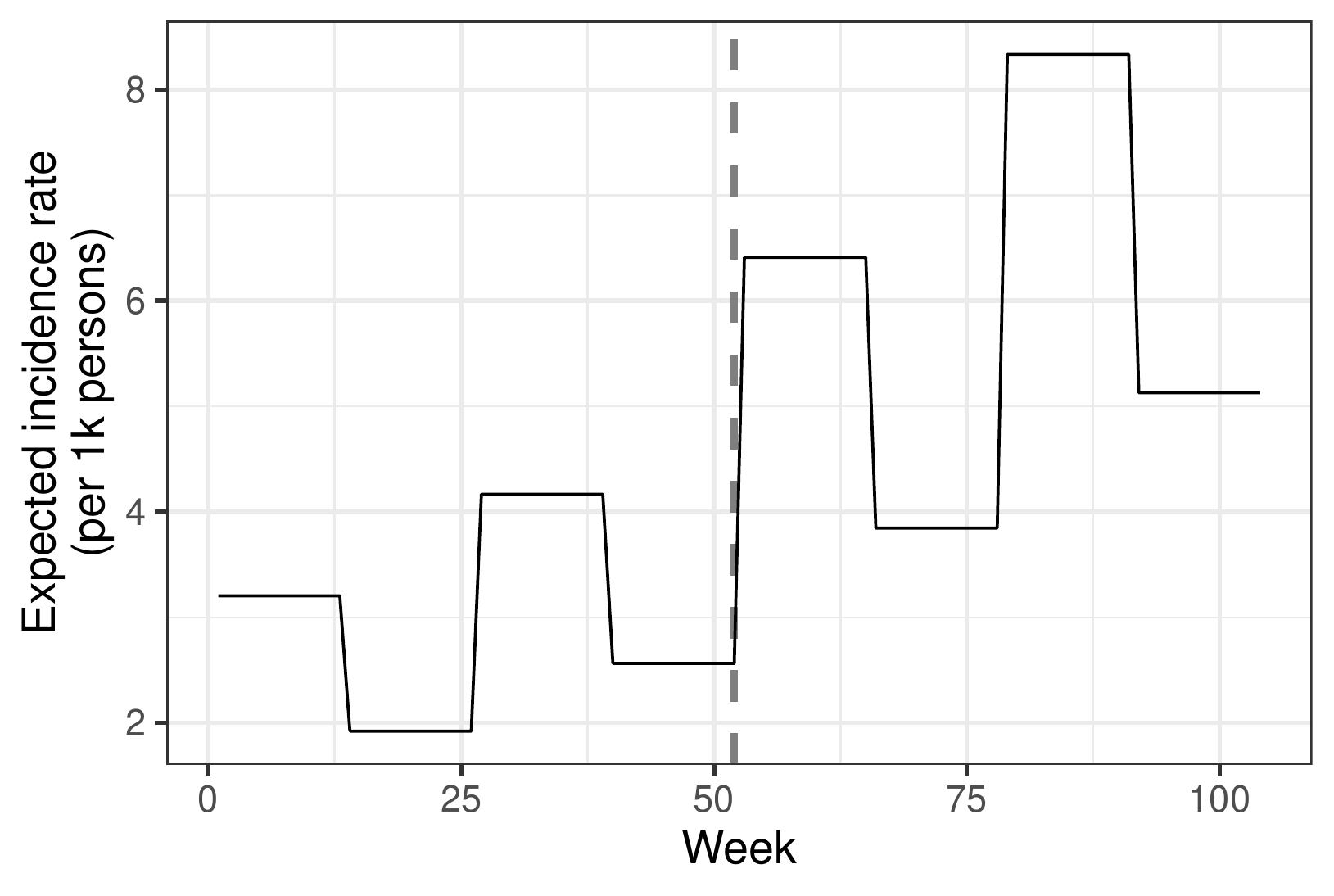}
        \caption{Time-varying background incidence rates of a hypothetical adverse event. Observations from the first year (week 1 to 52, left to gray vertical line) are used to calculate historical reference rates for the Historical Comparator design.}
    \end{subfigure}
    ~
    \begin{subfigure}{0.48\textwidth}
        \includegraphics[width=\textwidth]{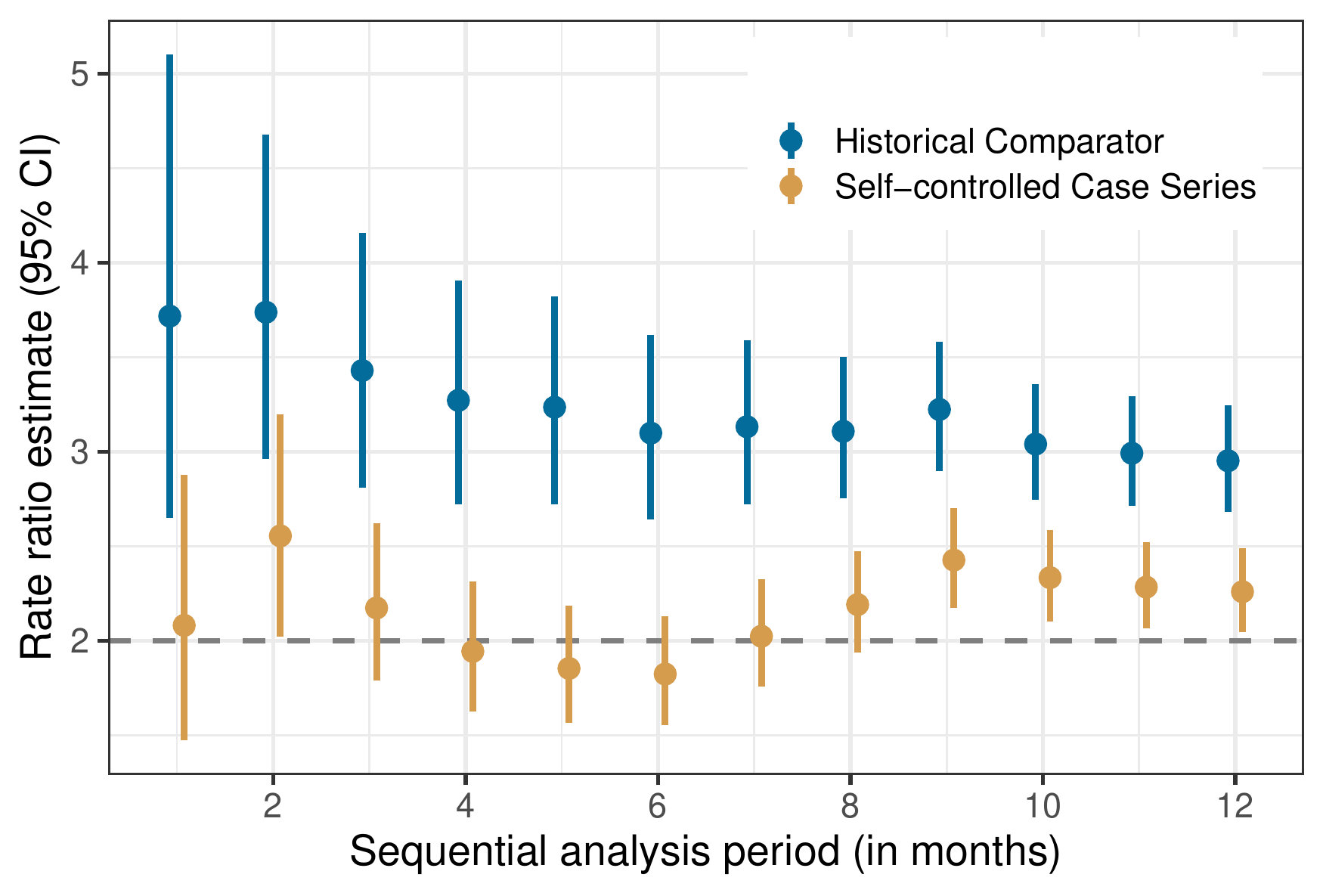}
        \caption{Rate ratio estimates (and 95\% confidence intervals) obtained in monthly sequential analyses using two common epidemiological designs. 
        The ground truth is $RR=2$, but both methods produce biased estimates. }
    \end{subfigure}
    \caption{\textbf{Two common epidemiological designs produce biased estimates due to residual systematic error}. Settings and results from a simulation experiment with $N=5000$ subject-level trajectories.}
    \label{fig:simulate-bias-issue}
\end{figure}

We simulate subject-level weekly incidence counts for the outcome of interest using the following Poisson count model: 
\begin{align*}
    Y_{ik} &\sim \text{Poisson}(\lambda_{ik}), \text{ where} \\
    \lambda_{ik} &= \exp[\alpha_k + x_i \gamma + \mathbbm{1}(i \text{ at risk in week }k) \times \beta].
\end{align*}
Here $Y_{ik}$ is the outcome incidence count for subject $i$ during week $k$, where $k = 1, 2, \ldots 104$ for which weeks $1$ to $52$ constitute the one-year historical period and weeks $53$ to $104$ constitute the present-time surveillance period. 
Further, $x_i$ is a randomly assigned binary covariate that represents individual $i$'s important characteristics, with coefficient $\gamma$. 
Intercept $\alpha_k$ denotes the expected weekly incidence rate without exposure, where there is a seasonality pattern as depicted in Figure~\ref{fig:simulate-bias-issue}(a).
The term $\mathbbm{1}(i \text{ at risk in week }k)$ is a binary indicator that takes value $1$ if subject $i$ is inside the risk window post vaccination, and $0$ otherwise. 
Recall that $\beta = \log(RR)$, with true value $\log(2)$ in this simulation. 
There is also a difference between the background incidence rates in the historical period (weeks 1 to 52) and in the surveillance period (weeks 53-104) where the average historical incidence rate captured in the data is only about 50\% of the present-time rate. 
Such differential background rates between historical and present times are commonly seen in real-world data, due to factors such as data collection bias, misclassification, time-varying population effects, shifts in diagnostics standards and guidelines, etc.


Subject-level event trajectories (including a one-shot vaccine exposure and possibly multiple occurrences of the outcome) are simulated for $N=5000$ individuals. We assume that each individual will be at risk of experiencing the adverse event within the first $6$ weeks following vaccine exposure. 
The vaccine uptake rate (number of people exposed to the hypothetical vaccine in each week) during week 53-104 has a seasonal pattern that takes the same shape as depicted in Figure~\ref{fig:simulate-bias-issue}(a). 

We implement the two epidemiological designs (and thus their implied statistical models) to estimate $RR$. Each design makes assumptions about the subject-level weekly incidence rate $\lambda_{ik}$ that are subject to residual systematic error in the data:

\vspace{-0.15in}
\paragraph{1. Historical comparator.} A Poisson regression model is used, where present-time incidence rates are compared against historical incidence rates to estimate $RR$, with seasonality effects within each year taken into account. That is,
\begin{equation*}
    \lambda_{ik} = \exp[s_k + x_i \gamma + \mathbbm{1}(i \text{ at risk in week }k) \times \beta],
\end{equation*}
where $s_k$ denotes the \emph{week of the year} for week $k$; that is, both week 2 and week 54 will have $s_k = 2$, and thus week 2 would serve as the historical reference period for week 54. 

\vspace{-0.15in}
\paragraph{2. SCCS.} A conditional Poisson regression model is used, where only ``cases'' (persons with at least one outcome during the surveillance period) are considered, and \emph{within each individual trajectory}, outcome incidents during at-risk times are compared against those not during at-risk times to estimate $RR$. 
That is,
\begin{equation*}
    \lambda_{ik} = \exp[\gamma_i + \mathbbm{1}(i \text{ at risk in week }k) \times \beta],
\end{equation*}
where $\gamma_i$ represents an individual factor specific to person $i$, thus effectively modeling each subject-level trajectory as an inhomogeneous Poisson process. 
For SCCS, only present-time data (weeks 53-104) are used.  

Analysis using each design is performed sequentially every month (i.e., every 4 or 5 weeks), on all data accrued up to the end of each month. For simplicity, we assume that there is no data accrual delay, such that events occurred in week $k$ are observed promptly and available for analysis. That is, the third analysis is run on \emph{all} data accumulated by the end of month 3. 

Figure~\ref{fig:simulate-bias-issue}(b) presents $RR$ estimates with 95\% confidence intervals obtained by each analysis design at each analysis timepoint (by each month). 
Here, we estimate $\beta$ as the parameter and then transform with $RR = \exp(\beta)$ to obtain point estimates and confidence intervals. 
The ground truth $RR=2$ is marked by the gray horizontal dashed line. 
Neither design produces satisfactorily accurate estimates. 
The historical comparator severely over-estimates $RR$, as it fails to capture the time-varying effect in that the background incidence rate during the surveillance period is higher than that during the historical period. 
Such temporal trends in observational data that are confounded with vaccine exposures during surveillance are  challenging to measure or model without knowing the true underlying data generative process. 
SCCS performs better, but its estimates can still be biased despite more data accrued over time. Even though SCCS can adjust for subject-level time-invariant covariates \citep{Whitaker2006,farrington2004control}, when the temporal patterns of vaccine exposures and outcome occurrences (and/or data accrual) are confounded, SCCS struggles to converge to the correct answer even with more data analyzed.  

The complexity of residual systematic error in observational data goes far beyond the unmeasured temporal effects or confounders simulated here to illustrate how these commonly applied epidemiological designs are subject to bias. 
The true mechanisms of systematic error may be a combination of selection bias, misclassification, unmeasured confounding and many other factors that are not directly observable, testable or adjustable through an epidemiological design alone. 
This motivates us to consider an approach that can diagnose and correct for bias induced by residual systematic error, within the context of sequential analysis where bias should be adjusted for adaptively over time.

\section{Methods}
\label{sec: methods}
In this section, we first provide a description of our statistical method for sequential analysis under a Bayesian framework, with joint statistical modeling to adaptively correct for bias induced by residual systematic error. 
Then we outline the design of acomprehensive empirical evaluation of the proposed method on multiple large-scale administrative health databases. 
The full protocol of the empirical evaluation study is publicly available at \url{https://suchard-group.github.io/Better/Protocol.html}.

\subsection{Bayesian sequential analysis with bias correction}

\label{sec: our-method}
We adopt a joint statistical modeling approach to sequentially estimate $\beta$ and adaptively correct for estimation bias induced by systematic error in observational data. 
The bias correction component of our model relies on estimating an empirical probabilistic distribution of the bias, accomplished with a fully data-driven approach. 

\paragraph{Overview of the Bayesian sequential analysis framework}
The input of our method includes the data $\bX_t$ accumulated up to each analysis timepoint $t$, a working statistical model that allows us to write down the joint density function $p(\bX_t \mid \beta)$ (as the likelihood function), and a \emph{prior distribution} $\pi_0(\beta)$ for $\beta$. 
We adopt a prior distribution such that the prior probabilities for $H_0$ and $H_1$ are both equal to $1/2$; one example prior choice to achieve this is to set $\pi_0(\beta)$ as a normal distribution with prior mean $0$. 

The output at each analysis time point $t$ is a \emph{posterior distribution} $\pi_t(\beta \mid \bX_t)$ for $\beta$, obtained using the Bayes rule as in \eqref{eq: belief-update}. 
In our framework we use the posterior probability for $H_1$, $P_{1,t}$ as the test statistic, defined in \eqref{eq: post-prob-calculation}. 
The posterior probability has been widely used for testing and decision-making, especially for Bayesian adaptive design \citep{cornfield1966bayesian,thall1995bayesian,smith2006implementation,zhou2008bayesian,berry2010bayesian,li2020bayesian}.
With a user-specified threshold $\delta_1$, we reject $H_0$ and thus declare a safety signal at the first time point $t$ when $P_{1,t} > \delta_1$. 
For example, if $\delta_1 = 0.95$, 
then as soon as the data evidence supports at least 95\% posterior credibility of $H_1$, we can stop the surveillance and claim that we have detected a safety signal. 

It is clear from this setup that our sequential analysis framework does \emph{not} require a pre-specified surveillance schedule as we do not pre-calculate a decision threshold, nor do we perform inference or make decisions based on the length or group sizes of a sequential study. 
Further, the posterior probability $P_{1,t}$ used for decision-making is naturally interpretable and evidence-driven, in that it quantifies how much we can trust a hypothesis based on the data accrued so far. 

\begin{figure}[ht]
    \centering
    \includegraphics[width=0.85\textwidth]{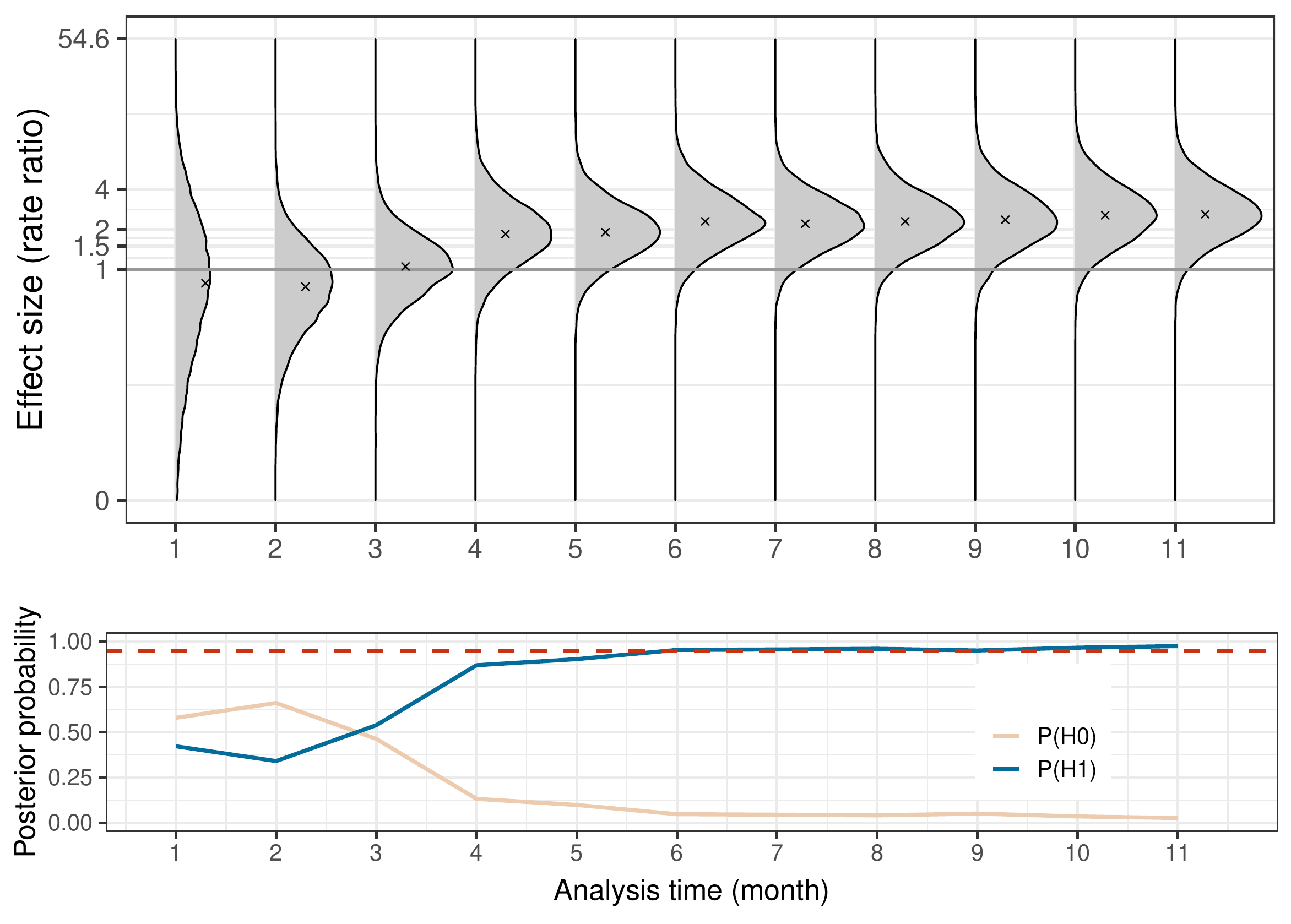}
    \caption{Example Bayesian sequential analysis for an exposure-outcome pair with $RR=2$. \textbf{Top}: posterior distributions $\pi_t(\beta \mid \bX_t)$learned with monthly sequential analyses; with more data accrued, the posterior distribution gets more concentrated around the ground truth. 
    \textbf{Bottom}: Posterior probability for $H_1$ (blue) and $H_0$ (yellow) with monthly sequential analyses, directly computed from posterior distributions; with threshold $\delta_1 = 95\%$, safety signal is declared when $P_{1,t} > \delta_1$ (by month 9 in this example). }
    \label{fig: posterior-updates}
\end{figure}

In Figure~\ref{fig: posterior-updates} we provide a graphical example of Bayesian sequential analysis for an exposure-outcome pair with true $RR=2$ (i.e., $\beta = \log(2)$), where analyses are performed on monthly accrued data. 
The shaded density curves in the top panel shows the posterior distributions $\pi_t(\beta \mid \bX_t)$ inferred from monthly sequential data with the posterior median marked by an ``x''. 
As more data are accrued over time, the posterior distribution becomes more concentrated around the true effect size value, as more data evidence would reduce uncertainty. 
By computing the area under the density curve for which $\beta > 0$ (or $\beta \leq 0$), we can easily update the posterior probability for $H_1$ (or $H_0$) at each analysis time point. 
The bottom panel shows the posterior probability values updated over time. 
Assuming a decision threshold $\delta_1 = 0.95$, we would reject $H_0$ and declare a safety signal as soon as $P(H_1 \mid \bX_t) > 0.95$ which happens at around month 9. 
This suggests that by month 9, we are 95\% certain that the alternative hypothesis (i.e., vaccination elevates risk of the adverse event) is true given the data evidence accrued. 

\paragraph{Adaptive bias correction} At each analysis time point, we adaptively correct for residual systematic error by learning an empirical distribution for the amount of bias and then effectively ``subtract'' bias in a probabilistic manner to produce a posterior distribution for a ``de-biased'' effect size. 

We do so by simultaneously analyzing a large set (typically between 50 and 100) of negative control outcomes. 
A negative control outcome is an outcome that is believed to have no significant association with a specific vaccine exposure \citep{Tchetgen2014control, Schuemie2014interpreting, schuemie2016robust}. 
Such outcomes are identified by the lack of evidence from reports, product labels and existing literature, and then confirmed by expert review. 
Intuitively, if the effect estimate of the vaccine exposure on a negative control outcome deviates from ``zero effect'', then the deviation provides an empirical characterization of the residual systematic error; such effect deviations for a large set of negative control outcomes can then be taken as a ``sample'' to inform the underlying bias in the effect-estimate of interest. 

For a specific vaccine exposure $e$ and a set of $M$ negative control outcomes, $\{o_i\}_{i=1}^M$, estimation of the true (unbiased) log-$RR$ $\beta_i$ related to negative control $o_i$ could be biased by a quantity of $b_i$. 
Assuming that the bias is additive on $\beta_i$, then the biased log-$RR$ quantity $\tilde\beta_i$ can be written as
\begin{equation*}
    \tilde\beta_i = \beta_i + b_i,
\end{equation*}
which, given the knowledge that $\beta_i = 0$ (since there is no association between $o_i$ and $e$), indicates
\begin{equation}
\label{eq: additive-bias}
    \tilde\beta_i = b_i.
\end{equation}
This suggests that profiling the estimated $\beta$ for all negative control outcomes produces an empirical distribution of the bias that is due to the residual systematic error in the observational data given a specific epidemiological design. 

More formally, assume that the biases $b_i$'s associated with the negative control outcomes $\{o_i\}_{i=1}^M$ are exchangeable and follow the same distribution:
\begin{equation}
    \label{eq: bias-model-normal}
    b_i \sim N(\bar{b}, \tau^2),
\end{equation}
where $\bar{b}$ denotes the unknown average bias and $\tau^2$ represents the unknown variability across different outcomes. 
The normal distribution could be replaced by any distribution with a density function; we illustrate our approach with the normal model here, but we have also implemented other distributions such as the $t$ distribution with various degrees of freedom. 
Parameters $\bar{b}$ and $\tau^2$ can be easily estimated by jointly fitting a normal or hierarchical normal model (or $t$ models), given either the estimates of $\tilde\beta_i$'s
or likelihood functions evaluated with the negative control outcomes. 

Naturally, this can be done in a dynamic and adaptive manner: at time $t$, we can update the estimated bias distribution using all the data related to negative control outcomes accrued up to time $t$. 
That is, suppose $q_t(b \mid \bX_t)$ represents the posterior predictive distribution for the bias $b$ learned using data accrued up to time $t$, and we can perform bias correction for the outcome of interest by re-writing \eqref{eq: belief-update} with regards to the bias effect $\tilde\beta$:
\begin{align}
    \tilde\pi_t(\tilde\beta, b \mid \bX_t) \propto \tilde{p}_t(\bX_t \mid \tilde\beta) \times \tilde\pi_0(\tilde\beta) \times q_t(b \mid \bX_t),
\end{align}
where $\tilde\pi_t$ denotes the joint posterior distribution for $\tilde\beta$ and $b$, $\tilde\pi_0$ is a prior distribution for $\tilde\beta$, and $\tilde{p}_t$ indicates the data likelihood function with regard to the biased effect. 
Using the relationship $\tilde\beta = \beta + b$ and thus $\beta = \tilde\beta - b$, inference of the true (unbiased) $\beta$ is straightforward from posterior samples of $\tilde\beta$ and $b$ via Markov chain Monte Carlo (MCMC). 

\begin{figure}
    \centering
    \includegraphics[width = \textwidth]{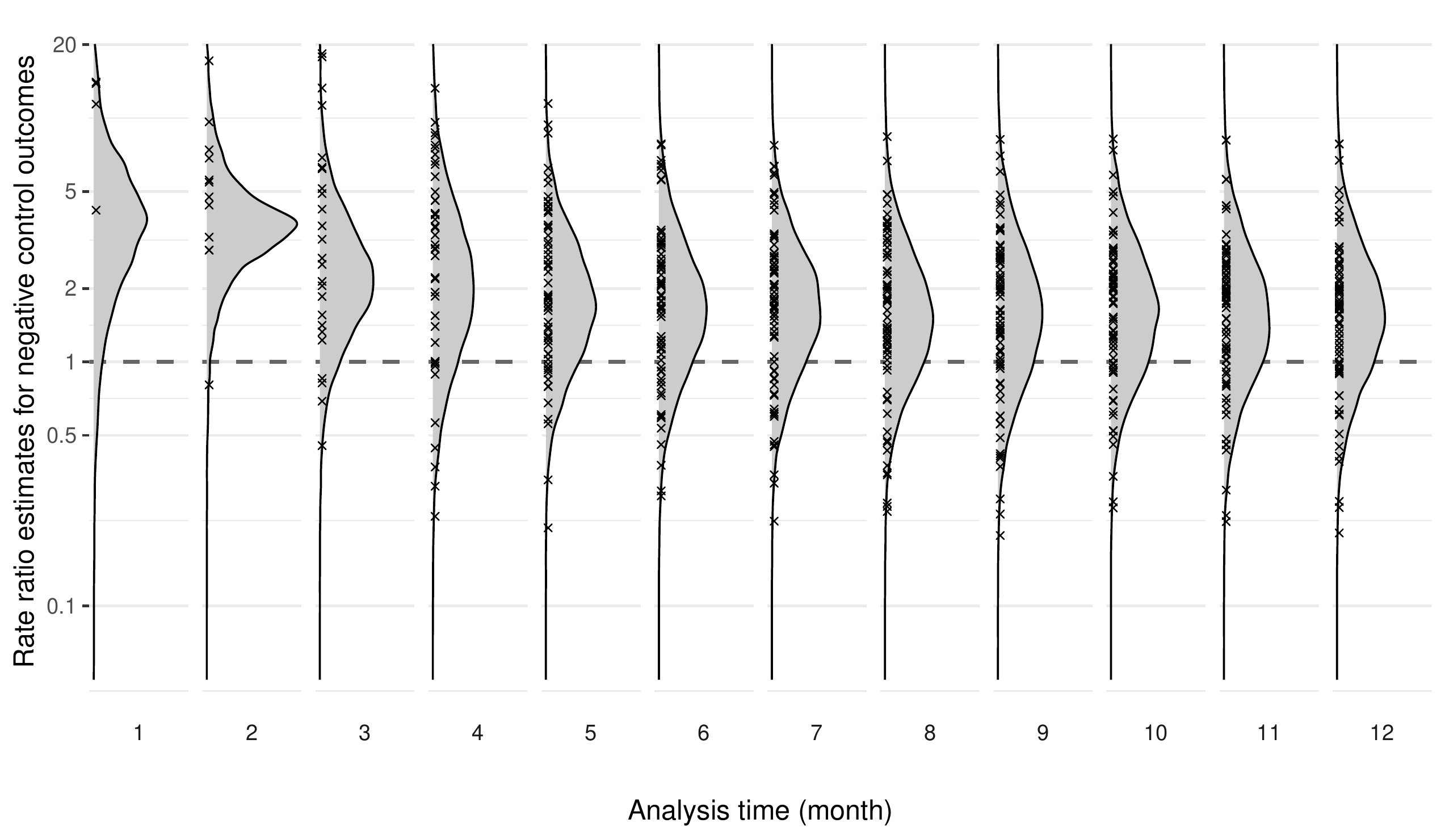}
    \caption{An example of learned bias distributions over time in sequential analysis. 
    Each ``x'' marks the maximum likelihood estimate (MLE) of log-$RR$ for each negative control outcome. The density curves characterize the empirical bias distribution learned from negative controls in monthly sequential analysis. 
    The ``zero-effect'' value $RR = 1$ is marked by the gray dashed line; there is a positive bias since the majority of densities under these curves lie above the $RR=1$ line.}
    \label{fig:bias-distribution-over-time}
\end{figure}

In Figure~\ref{fig:bias-distribution-over-time} we present a graphical example of empirical bias distributions $q_t(b \mid \bX_t)$ learned sequentially from monthly accrued data. 
The density curve by month $t$ characterizes the learned bias distribution through negative control analysis up to month $t$, where each ``x'' marks the maximum likelihood estimate (MLE) of $\tilde\beta$ for each negative control outcome. 
Over time, more negative control estimates become available (note that more ``x''s are present for later months) and the empirical bias distribution would also stabilize. 
In this example, a positive bias seems to persist, as the majority of the density lies above $RR = 1$ (the gray dashed line representing a null effect). 
We include more details of the bias correction procedure in the Supporting Materials. 
Furthermore, 
we provide a pseudocode-style sketch of the sequential analysis procedure with adaptive bias correction in Algorithm~\ref{alg:Bayesian-framework-algorithm} and a complete, open-source implementation in the R package \texttt{EvidenceSynthesis} available at \url{https://github.com/OHDSI/EvidenceSynthesis}. 

\newcommand{\nc}{\text{\tiny NC}}
\newcommand{\ooi}{\text{\tiny OOI}}

\begin{algorithm}
\caption{Bayesian bias correction (BBC) for sequential analysis}
\label{alg:Bayesian-framework-algorithm}
\hspace*{\algorithmicindent} \textbf{Input}: sequential data $\bX_t$, working model likelihood $\tilde{p}$, prior $\tilde\pi_0$, threshold $\delta_1$ (optional)\\
\hspace*{\algorithmicindent} \textbf{Output}: posterior distributions $\pi_t$, study stopping time $\tau_s$ (optional)
\setstretch{1.2}
\begin{algorithmic}[1]

\Procedure{BBC}{}       
\For{$t = 1, 2, \ldots, T_{\max}$} \Comment{Sequentially analyze data}
\State Obtain/collect accumulated data $\bX_t$ up to time $t$
\State Split $\bX_t$ into $\bX_t^{(\nc)}$ (data on negative controls) and $\bX_t^{(\ooi)}$ (data on outcome of interest)
\State Learn empirical bias distribution $q_t(b \mid \bX_t^{(\nc)})$
\State Obtain posterior for biased log-$RR$ via $\tilde\pi_t(\tilde\beta \mid \bX_t^{(\ooi)}) \propto \tilde{p}_t(\bX_t \mid \tilde\beta) \times \tilde\pi_0(\tilde\beta)$
\For{$s = 1, 2, \ldots, S$}   \Comment{Learn unbiased $\beta$ via MCMC}
\State sample $b^{(s)} \sim q_t(b \mid \bX_t^{(\nc)})$
\State sample $\tilde\beta^{(s)} \sim \tilde\pi_t(\tilde\beta \mid \bX_t^{(\ooi)})$
\State $\beta^{(s)} \gets \tilde\beta^{(s)} - b^{(s)}$ \Comment{Use relationship $\beta = \tilde\beta - b$}
\EndFor
\State Compute $P_{1,t}$, posterior probability of $H_1$ via
\begin{equation}
    \hat{P}_{1,t} := \frac{1}{S}\sum_{s=1}^S\mathbbm{1}(\beta^{(s)} > 0)
\end{equation}
\Comment{Counting fraction of MCMC samples with $\beta > 0$}
\If{$\hat{P}_{1,t} > \delta_1$} \Comment{Stop if decision threshold crossed; optional}
    \State Exit \textbf{for} loop, $\tau_s \gets t$
\ElsIf{$\bX_t$ exhausts all available data}
    \State Exit \textbf{for} loop
\EndIf
\EndFor 
\EndProcedure
\end{algorithmic}
\end{algorithm}

\subsection{Empirical evaluation on large-scale observational healthcare databases}

The over-arching goal of our empirical evaluation is to evaluate the performance of the proposed Bayesian sequential analysis framework on real-world observational health databases, and benchmark against MaxSPRT. 
We use six historical vaccines with known side effects, a large set of experimental control outcomes for which the true $RR$ values are known, two commonly used analysis designs (the historical comparator and SCCS), and run the analyses on a network of observational healthcare databases mapped to Observational Health Data Science and Informatics (OHDSI, \cite{hripcsak2015observational}) collaborative's Observational Medical Outcomes Partnership (OMOP) common data model (CDM) \citep{overhage2012validation}.
We simulate the data accrual process by making sequential batches of data available for analysis at one-month intervals, where subject-level records become available in chronological order. 
We evaluate the proposed Bayesian bias correction (BBC) framework and MaxSPRT on testing-based metrics (e.g., Type 1 error rates and statistical power) as well as estimation-based metrics (e.g., mean squared errors). 

Below we provide a brief overview of each aspect of the design of the empirical evaluation:
%
\paragraph{(a) Exposure-outcome pairs}
We use six existing vaccines (or vaccine groups) including seasonal influenza, pandemic influenza (H1N1pdm), human papillomavirus (HPV), and varicella zoster virus, with data collected during specific time periods, as shown in Table~\ref{tab: exposures}. 
The starting point and end point of time periods used for the study and historical reference are recorded by ``Start Date'' and ``End Date'' under the ``Study'' and ``Historical'' columns, respectively. 

Since the zoster (Shingrix) and HPV (Gardasil 9) vaccines have two doses, we spilt the two doses into two separate exposures and also consider a third exposure defined as receiving either dose. 
For instance, for the zoster (Shingrix) vaccine, we consider three exposures: zoster first dose, zoster second dose, and zoster first or second dose, where all of them share the same observational periods for the study and historical reference. 
Therefore, in total, we consider $10$ vaccine exposures. 
The codes and inclusion-exclusion criteria for constructing these exposure cohorts can be found in the Appendix of the online study protocol at \url{https://suchard-group.github.io/Better/Protocol.html#Appendix}. 

{
\small
\begin{table}
\caption{Vaccine exposures of interest, with start and end dates of data used for safety surveillance, and start and end dates of data used for historical reference.}
\label{tab: exposures}
\centering
\begin{tabular}{lllll}
  \toprule 
  & \multicolumn{2}{c}{\textbf{Study}} & \multicolumn{2}{c}{\textbf{Historical}} \\[.2em]
\textbf{Vaccine Exposure Name }
& \multicolumn{1}{c}{\textbf{Start Date}} & \multicolumn{1}{c}{\textbf{End Date}} 
& \multicolumn{1}{c}{\textbf{Start Date}} & \multicolumn{1}{c}{\textbf{End Date}} \\ 
  \midrule
H1N1pdm & 09/01/2009 & 05/31/2010 & 09/01/2008 & 05/31/2009 \\ 
  Seasonal flu (Fluvirin) & 09/01/2017 & 05/31/2018 & 09/01/2016 & 05/31/2017 \\ 
  Seasonal flu (Fluzone) & 09/01/2017 & 05/31/2018 & 09/01/2016 & 05/31/2017 \\ 
  Seasonal flu (All) & 09/01/2017 & 05/31/2018 & 09/01/2016 & 05/31/2017 \\ 
  Zoster (Shingrix) & 01/01/2018 & 12/31/2018 & 01/01/2017 & 12/31/2017 \\ 
  HPV (Gardasil 9) & 01/01/2018 & 12/31/2018 & 01/01/2017 & 12/31/2017 \\ 
   \bottomrule
\end{tabular}
\end{table}
}

We use a large set of experimental control outcomes including negative control outcomes and positive control outcomes that share the systematic error structures of the negative controls. 
We select a single set of 93 negative control outcomes for all six vaccine groups that match the severity and prevalence of suspected vaccine adverse effects.
We first generated a longer candidate list of negative controls based on similarity of prevalence and percent of diagnoses that were recorded in an inpatient setting (as a proxy for severity) and then finalized the list with manual review by clinical experts. 
We use a large set of negative controls in order to represent a wide range of diseases or conditions that can cover a broad population of potential systematic errors and sources of bias. 
Some example negative control outcomes we have identified include chronic pancreatitis, hypothermia and leukemia.

We synthesize positive control outcomes from the negative control outcomes so that we have knowledge about the true effect sizes of vaccine exposures to those positive controls. 
We choose to use these synthesized positive control outcomes instead of real positive outcomes, as real positive outcomes are problematic for a multitude of reasons \citep{Schuemie2018-zi}. 
First, adverse effects of vaccines are rarely well established, and even for an established effect, the effect size (or magnitude) is never known with absolute certainty or precision. 
Second, for a well known adverse effect, regulatory actions (such as restriction of use or careful monitoring) are often taken to ameliorate the risk, which will then mask such effect in real-world data. 
In our study, suppose the likelihood function regarding $\beta$ evaluated with a particular negative control outcome is denoted by $f_{\text{nc}}(\beta)$. 
We synthesize a positive control outcome by directly synthesizing its likelihood function through \emph{horizontally} moving $f_{\text{nc}}(\beta)$ to the positive direction by a desired amount.
For instance, for a positive control outcome with true $\beta = \log(2)$, its synthesized likelihood function is then $f_{pc,\text{rr=2}}(\beta^*)  = f_{nc}(\beta - 2)$.
In our empirical evaluation, for each negative control outcome, we synthesize three positive control outcomes, with $RR = 1.5$, $2$ and $4$ (i.e., $\beta = \log(1.5)$, $\log(2)$ and $\log(4)$), respectively. 

We also investigate Guillain-Barré syndrome (GBS) as a special outcome of interest. 
Previous studies have found a significant association of GBS with the zoster (Shingrix) vaccination \citep{goud2021risk} with an estimated $RR = 2.84$. 
As a brief case study (Section~\ref{sec: case-study}), we return to this association using the same epidemiological design and a similar data source to examine the findings made by the proposed BBC framework and by MaxSPRT.

\paragraph{(b) Data sources} 

Our evaluation uses the following US observational healthcare databases that have been widely used in previous OHDSI methodological and clinical studies (recent examples include \cite{schuemie2022vaccine,ostropolets2022covid,suchard2019comprehensive,hripcsak2016characterizing}): 

\begin{enumerate}
    \item IBM MarketScan Commercial Claims and Encounters \textbf{(CCAE)}: adjudicated health insurance claims (e.g., inpatient, outpatient, and outpatient pharmacy) from large employers and health plans who provide private healthcare coverage to employees, their spouses and dependents. Population size $\approx$ 142 million. 
    \item IBM MarketScan Medicare Supplemental Database \textbf{(MDCR)}: adjudicated health insurance claims of retirees with primary or Medicare supplemental coverage through privately insured fee-for-service, point-of-service or capitated health plans. Population size $\approx$ 10 million. 
    \item IBM MarketScan Multi-State Medicaid Database \textbf{(MDCD)}: contains adjudicated health insurance claims for Medicaid enrollees from multiple states and includes hospital discharge diagnoses, outpatient diagnoses and procedures, and outpatient pharmacy claims. Population size $\approx$ 26 million. 
    \item Optum Clinformatics Data Mart \textbf{(Clinformatics)}: inpatient and outpatient healthcare insurance claims for enrollees insured by commercial insurances or Medicare. Population size $\approx$ 85 million. 
    \item Optum\textsuperscript{®} de-identified Electronic Health Record dataset \textbf{(Optum EHR)}: EHRs containing clinical information, prescriptions, lab results, vital signs, body measurements, diagnoses and procedures derived from clinical notes from both inpatient and outpatient environments using natural language processing. Population size $\approx$ 93 million. 
    \item The Columbia University Irving Medical Center database \textbf{(CUIMC)}: EHR and administrative databases including inpatient and outpatient records on demographics, visits, drugs, devices, measurements and other observations (e.g., symptoms), collected from CUIMC's primary care practices in northern Manhattan and surrounding areas. Population size $\approx$ 6.7 million. 
\end{enumerate}

We provide more details on these databases in the online study protocol at \url{https://suchard-group.github.io/Better/Protocol.html#82_Data_sources}. All databases have been mapped to the OMOP CDM \citep{overhage2012validation}. 
OHDSI's Data Quality Dashboard \citep{blacketer2021increasing} has been used to assess and ensure data quality prior to processing any analysis results.

\paragraph{(c) Study designs and statistical models}
We adopt two commonly used epidemiological designs to construct counterfactuals (or controls) and compare them against exposed person-times in order to evaluate the relation between an outcome and a vaccine exposure. 
For each design, we consider two specifications about ``time-at-risk'', i.e., the time window during which a subject is at risk of experiencing an adverse event due to the vaccine exposure: 
(1) 1-28 days post vaccination, or (2) 1-42 days post vaccination.

\textbf{1. Historical comparator.} This is a conventional design commonly used in safety surveillance \citep{lieu2007real,black2009importance,yih2009assessment,belongia2010real,yih2011active,buttery2011intussusception,wijnans2013incidence,barker2014pandemic,li2021characterising,klein2021surveillance}. 
Upon computing a historical incidence rate for an adverse event during some historical time period, this design constructs the counterfactual by estimating the expected incidence rate for the present time period given the historical rate.
We implement four variants with varying choices of historical time periods and covariate adjustments: 
\begin{itemize}
    \item \textbf{Unadjusted, entire year}. Using a single rate computed across the entire historical year for the entire population.
    \item \textbf{Age and gender adjusted, entire year}. Using a rate stratified by age (in 10 year increments) and gender, computed across the entire historical year. This allows the expected rate to be adjusted for the demographics of the vaccinated.
    \item \textbf{Unadjusted, time-at-risk relative to outpatient visit}. Using a single rate computed during the time-at-risk relative to a random outpatient visit in the historical year.
    \item \textbf{Age and gender adjusted, time-at-risk relative to outpatient visit}. Using a rate stratified by age and gender, computed during the time-at-risk relative to a random outpatient visit in the historical year.
\end{itemize}

The historical comparator design implies a Poisson count model for the adverse event counts, where the Poisson rate parameter is the incidence rate.
Therefore, we use a Poisson likelihood function under this design. 

\textbf{2. SCCS.} This is a more recently developed design that constructs counterfactual time periods for a subject who has experienced the adverse outcome (a ``case'') using their own trajectory \citep{farrington1995new,
Whitaker2006,farrington2004control,salmon2013association,grave2020seasonal}. Similarly, we implement five variants given different choices of control time periods and covariate adjustments:
\begin{itemize}
    \item \textbf{Unadjusted SCCS excluding pre-vaccination window}. A simple SCCS, using all patient time when not at risk as the control time, with the exception of the 30 days prior to vaccination which is excluded from the analysis to avoid bias due to contra-indications.
    \item  \textbf{Age \& season adjusted SCCS excluding pre-vaccination window.} An SCCS adjusting for age and season, also excluding the 30 days prior to vaccination. Age and season will be modeled to be constant within each calendar month, and vary across months as bicubic splines.
    \item \textbf{Unadjusted SCCS excluding all pre-vaccination time}. A simple SCCS discarding all time prior to vaccination and only using post vaccination time as control time periods.
    \item \textbf{Self-controlled risk interval (SCRI) variant with prior control interval.} An SCRI using a control interval of 43 to 15 days prior to vaccination.
    \item \textbf{SCRI with post control interval}.  An SCRI using a control interval of 43 to 71 days after to vaccination.
\end{itemize}

The SCCS or SCRI design implies a conditional Poisson model for the adverse event outcomes, where the incidence count during each time interval follows a (conditional) Poisson distribution with a rate parameter specific to the instantaneous risk inside the interval. 
Therefore, we use a conditional Poisson (or Poisson process) likelihood function under this design. 


\paragraph{(d) Bayesian analysis choices} 
The Bayesian sequential analysis procedure requires the user to specify two inputs: the prior distribution $\pi_0$, and the decision threshold $\delta_1$ for posterior probability $P_{1,t}$. 
We consider three choices for the prior distribution $\pi_0$ for $\beta$, all as normal distributions with mean $\mu_0 = 0$ but with different variances $\sigma_0^2$:
\begin{itemize}
    \item \textbf{Conservative prior} with $\sigma_0^2 = 1.5$.
    \item \textbf{Moderately informed prior} with $\sigma_0^2 = 4$.
    \item \textbf{Diffuse prior} with $\sigma_0^2 = 10$.
\end{itemize}
We note that the ``diffuse prior'' leads to inference results that are close to maximum-likelihood estimates (MLEs) under frequentist inference. 

We also consider three choices for the posterior probability threshold $\delta_1$: $0.8, 0.9$ and $0.95$. 
However, since the choice of $\delta_1$ has no impact on posterior estimation (but only on decision making), this threshold can also be flexibly chosen and adjusted after obtaining all inference results at the end of the surveillance period, given the retrospective nature of our evaluation study. 

\paragraph{(e) Evaluation metrics}
We benchmark the proposed BBC framework against MaxSPRT using a set of metrics to evaluate both testing and estimation performance. Note that we compute all the metrics for each framework, each design variant, each choice of risk window, and every combination of Bayesian analysis choices across all databases and vaccine exposures. 

Testing-oriented metrics include:
\begin{itemize}
    \item \textbf{Type 1 error rate (false positive rate).} Estimated as the fraction of negative controls for which a safety signal declares itself (testing statistic exceeding the threshold). 
    \item \textbf{Type 2 error rate (false negative rate).} Estimated as the fraction of positive controls for which a safety signal is \emph{not} declared, stratified by effect sizes of the 
    positive control outcomes. 
    \item \textbf{Sensitivity and specificity.} Sensitivity is equivalent to statistical power, which is $1-$ Type 2 error rate. 
    Specificity is defined as $1-$ Type 1 error rate. 
    \item \textbf{Time-to-detection}. The number of analyses (months) until signals are declared for a specified fraction ($25\%$ or $50\%$) of positive controls, stratified by effect sizes. 
\end{itemize}
Given the sequential nature of the analysis, we report all testing-oriented metrics (except time to detection)  measured over time. 

Estimation-oriented metrics include:
\begin{itemize}
    \item \textbf{Mean squared error (MSE)}. Mean squared error between the point estimate of $\beta$ and the true $\beta$.
    \item \textbf{Coverage rate.} The fraction of 95\% confidence or Bayesian credible intervals that cover the true $\beta$, stratified by true effect sizes of the negative or positive control outcomes. 
    \item \textbf{Non-estimable rate.} The fraction of control outcomes for which an estimate cannot be produced. 
\end{itemize}

\paragraph{(f) Data characteristics overview}
We further present some key data characteristics in Table~\ref{tab:data-characteristics}. 
Here we summarize, for each vaccine exposure and each database, the total number of subjects exposed, the total accumulated exposure time (in person-years), and some summary statistics (median and inter-quartile range) of the incident count and incident rate across all negative control outcomes, with the ``time-at-risk'' taken as 1-28 days post vaccination. More data characteristics information is provided in the Supporting Materials. 

\clearpage{
\footnotesize
\begin{longtable}{lrrr@{\ }lr@{\ }l}
\label{tab:data-characteristics}\\ 
\caption{\textbf{Data characteristics by exposures and databases.} Column ``Exposure Subjects'' shows the total number of unique people with vaccination exposure considered in the analysis. 
Column ``Exposure Days'' shows the cumulative at-risk days for all exposure subjects. 
Column ``Outcome counts'' shows the median event counts across 93 negative control outcomes during all exposure time periods; numbers in the parentheses are the 25th and 75th percentiles (i.e., the interquartile range, IQR). 
Column ``Incidence rates'' shows the incidence rate per person-year (incident count divided by exposure person-years) across all 93 negative control outcomes, similarly with the median and IQR. 
Summary is presented for design choices with 1-28 days after vaccine exposure considered as the ``time-at-risk''. }\\
    \vspace{.03in}\\
    \toprule 
   \multirow{2}{*}{\textbf{Database}} & \textbf{Exposure} & \textbf{Exposure} & \multicolumn{2}{c}{\textbf{Outcome counts}} & \multicolumn{2}{c}{\textbf{Incidence rates} ($\times 10^{-4}/$ p-yrs)} \\
    & \textbf{Subjects} & \textbf{Person-years} & \textbf{median} & \multicolumn{1}{l}{\textbf{IQR}} & \textbf{median} & \multicolumn{1}{l}{\textbf{IQR}} \\
       \midrule 
    \multicolumn{5}{l}{\textbf{H1N1 vaccination}} \\ [0.25em] 
  CCAE &   753,592 &  56,359.96 & 12.0 & [6.0, 42.0] & 2.13 & [1.06, 7.45] \\ 
  CUIMC &    10,611 &     784.28 & 4.0 & [2.0, 9.0] & 51.00 & [25.50, 114.76] \\ 
  MDCD &   206,865 &  15,447.37 & 4.0 & [2.0, 13.5] & 2.59 & [1.29, 8.74] \\ 
  MDCR &    12,913 &     975.82 & 2.0 & [1.0, 4.0] & 20.50 & [10.25, 40.99] \\ 
  Clinformatics &   457,565 &  34,373.75 & 13.0 & [5.0, 35.0] & 3.78 & [1.45, 10.18] \\ 
  Optum EHR &   156,467 &  11,962.63 & 5.0 & [2.0, 11.0] & 4.18 & [1.67, 9.20] \\ 
  \midrule \multicolumn{5}{l}{\textbf{Seasonal flu vaccination (Fluvirin)}} \\ [0.25em] 
  CCAE &   119,186 &   9,022.36 & 4.0 & [2.0, 11.0] & 4.43 & [2.22, 12.19] \\ 
  CUIMC &       230 &      17.39 & 1.0 & [1.0, 1.0] & 575.20 & [575.20, 575.20] \\ 
  MDCD &    15,282 &   1,160.04 & 2.0 & [1.0, 6.2] & 17.24 & [8.62, 53.88] \\ 
  MDCR &       822 &      61.75 & 1.0 & [1.0, 1.0] & 161.93 & [161.93, 161.93] \\ 
  Clinformatics &   189,184 &  14,372.50 & 7.0 & [3.0, 14.8] & 4.87 & [2.09, 10.26] \\ 
  Optum EHR &    14,706 &   1,120.67 & 2.0 & [1.0, 4.0] & 17.85 & [8.92, 35.69] \\ 
  \midrule \multicolumn{5}{l}{\textbf{Seasonal flu vaccination (Fluzone)}} \\ [0.25em] 
  CCAE &       957 &      69.60 & 1.0 & [1.0, 1.0] & 143.67 & [143.67, 143.67] \\ 
  CUIMC &     3,397 &     260.37 & 1.0 & [1.0, 3.5] & 38.41 & [38.41, 134.42] \\ 
  MDCD &     3,357 &     256.48 & 2.0 & [1.0, 2.0] & 77.98 & [38.99, 77.98] \\ 
  MDCR &    34,414 &   2,615.74 & 2.0 & [1.0, 5.5] & 7.65 & [3.82, 21.03] \\ 
  Clinformatics &   798,816 &  61,118.75 & 40.5 & [12.0, 96.8] & 6.63 & [1.96, 15.83] \\ 
  Optum EHR &   337,218 &  23,690.50 & 12.5 & [5.0, 28.0] & 5.28 & [2.11, 11.82] \\ 
  \midrule \multicolumn{5}{l}{\textbf{Seasonal flu vaccination (All)}} \\ [0.25em] 
  CCAE & 3,516,811 & 266,796.18 & 78.5 & [27.5, 214.8] & 2.94 & [1.03, 8.05] \\ 
  CUIMC &   117,339 &   8,939.38 & 12.0 & [5.0, 39.0] & 13.42 & [5.59, 43.63] \\ 
  MDCD & 1,237,934 &  94,034.65 & 40.0 & [19.0, 136.0] & 4.25 & [2.02, 14.46] \\ 
  MDCR &   264,636 &  20,078.82 & 18.0 & [7.8, 43.5] & 8.96 & [3.86, 21.66] \\ 
  Clinformatics & 3,399,471 & 258,963.54 & 166.0 & [52.8, 330.8] & 6.41 & [2.04, 12.77] \\ 
  Optum EHR & 2,536,334 & 190,273.53 & 100.0 & [42.0, 251.0] & 5.26 & [2.21, 13.19] \\ 
  \midrule \multicolumn{5}{l}{\textbf{First HPV vaccination (Gardasil 9)}} \\ [0.25em] 
  CCAE &   376,341 &  28,202.88 & 6.0 & [2.0, 21.0] & 2.13 & [0.71, 7.45] \\ 
  CUIMC &     7,670 &     574.87 & 1.0 & [1.0, 2.5] & 17.40 & [17.40, 43.49] \\ 
  MDCD &   236,683 &  17,767.07 & 4.0 & [1.0, 21.0] & 2.25 & [0.56, 11.82] \\ 
  Clinformatics &   173,228 &  12,938.73 & 6.0 & [1.2, 9.0] & 4.64 & [0.97, 6.96] \\ 
  Optum EHR &   233,985 &  17,301.64 & 5.0 & [2.0, 15.5] & 2.89 & [1.16, 8.96] \\ 
  MDCR & 0 & 0 & \\ 
  \midrule \multicolumn{5}{l}{\textbf{Second HPV vaccination (Gardasil 9)}} \\ [0.25em] 
  CCAE &    49,283 &   3,486.95 & 2.0 & [1.0, 4.0] & 5.74 & [2.87, 11.47] \\ 
  CUIMC &     1,172 &      84.26 & 1.0 & [1.0, 1.5] & 118.68 & [118.68, 178.03] \\ 
  MDCD &    15,065 &   1,066.11 & 2.0 & [1.2, 4.0] & 18.76 & [11.72, 37.52] \\ 
  Clinformatics &    21,377 &   1,508.51 & 2.0 & [1.0, 4.0] & 13.26 & [6.63, 26.52] \\ 
  Optum EHR &    28,336 &   2,005.88 & 2.0 & [1.0, 3.0] & 9.97 & [4.99, 14.96] \\ 
  MDCR & 0 & 0 & \\ 
  \midrule \multicolumn{5}{l}{\textbf{First or second HPV vaccination (Gardasil 9)}} \\ [0.25em] 
  CCAE &   378,052 &  31,689.85 & 6.0 & [2.0, 22.5] & 1.89 & [0.63, 7.10] \\ 
  CUIMC &     7,726 &     659.14 & 1.0 & [1.0, 2.5] & 15.17 & [15.17, 37.93] \\ 
  MDCD &   237,455 &  18,833.17 & 4.0 & [1.0, 21.8] & 2.12 & [0.53, 11.55] \\ 
  Clinformatics &   174,692 &  14,447.25 & 6.0 & [1.8, 10.2] & 4.15 & [1.21, 7.09] \\ 
  Optum EHR &   234,518 &  19,307.52 & 5.5 & [2.0, 16.8] & 2.85 & [1.04, 8.68] \\ 
  MDCR & 0 & 0 & \\ 
  \midrule \multicolumn{5}{l}{\textbf{First zoster vaccination (Shingrix)}} \\ [0.25em] 
  CCAE &   148,190 &  11,004.20 & 4.0 & [2.0, 13.0] & 3.63 & [1.82, 11.81] \\ 
  CUIMC &    11,182 &     835.11 & 2.0 & [1.0, 4.5] & 23.95 & [11.97, 53.88] \\ 
  MDCD &    11,407 &     851.97 & 2.0 & [1.0, 6.0] & 23.48 & [11.74, 70.43] \\ 
  MDCR &    52,789 &   3,952.34 & 4.0 & [2.0, 9.8] & 10.12 & [5.06, 24.67] \\ 
  Clinformatics &   229,463 &  17,113.06 & 10.0 & [5.0, 27.0] & 5.84 & [2.92, 15.78] \\ 
  Optum EHR &   219,665 &  16,251.16 & 8.0 & [3.0, 27.0] & 4.92 & [1.85, 16.61] \\ 
  \midrule \multicolumn{5}{l}{\textbf{Second zoster vaccination (Shingrix)}} \\ [0.25em] 
  CCAE &    72,063 &   5,117.29 & 3.0 & [1.5, 7.5] & 5.86 & [2.93, 14.66] \\ 
  CUIMC &     4,229 &     307.32 & 1.5 & [1.0, 2.8] & 48.81 & [32.54, 89.48] \\ 
  MDCD &     5,379 &     388.03 & 1.0 & [1.0, 3.0] & 25.77 & [25.77, 77.31] \\ 
  MDCR &    30,218 &   2,161.22 & 3.0 & [1.0, 7.0] & 13.88 & [4.63, 32.39] \\ 
  Clinformatics &   119,556 &   8,506.70 & 6.0 & [2.0, 13.0] & 7.05 & [2.35, 15.28] \\ 
  Optum EHR &    63,464 &   4,585.85 & 4.0 & [2.0, 9.0] & 8.72 & [4.36, 19.63] \\ 
  \midrule \multicolumn{5}{l}{\textbf{First or second zoster vaccination (Shingrix)}} \\ [0.25em] 
  CCAE &   149,219 &  16,121.49 & 6.0 & [3.0, 13.8] & 3.72 & [1.86, 8.53] \\ 
  CUIMC &    11,211 &   1,142.45 & 2.0 & [1.0, 6.0] & 17.51 & [8.75, 52.52] \\ 
  MDCD &    11,556 &   1,239.99 & 2.0 & [2.0, 7.0] & 16.13 & [16.13, 56.45] \\ 
  MDCR &    53,384 &   6,113.56 & 5.0 & [2.0, 15.0] & 8.18 & [3.27, 24.54] \\ 
  Clinformatics &   232,669 &  25,619.78 & 13.0 & [5.0, 36.0] & 5.07 & [1.95, 14.05] \\ 
  Optum EHR &   220,106 &  20,837.00 & 10.0 & [4.0, 32.5] & 4.80 & [1.92, 15.60] \\
   \bottomrule
   \end{longtable}
   }

\section{Evaluation results}
\label{sec: results}

In this section we discuss results from the empirical evaluation comparing the performance of our proposed sequential analysis framework with Bayesian bias correction (BBC) and that of MaxSPRT. 
All MaxSPRT analyses are implemented with significance level $\alpha = 0.05$, and with a pre-specified surveillance schedule that \emph{exactly matches} the actual data accrual process --- this is technically impossible in practice, and such a choice is favorable to MaxSPRT and should produce the best possible performance of using MaxSPRT for safety surveillance on real-world data.

Since there are more than 7,000,000 sets of analysis results from this large-scale empirical evaluation, we present a collection of representative results in terms of the test-oriented metrics, 
and provide summary statistics in terms of the 
estimation-oriented metrics. If not otherwise specified, in this section we focus on Bayesian analyses using the ``moderately informed prior'' with prior variance $\sigma_0^2 = 4$. 
We choose not to report on studies with insufficient evidence where the maximum incidence count across all negative control outcomes is lower than $2$, 
after examining the data characteristics (Table~\ref{tab:data-characteristics}) but before inspecting the results. 
For completeness, we make all results publicly available through an R ShinyApp at \url{https://data.ohdsi.org/BetterExplorer}.

\subsection{The Bayesian framework controls Type 1 error better}

\begin{figure}
\centering
\begin{subfigure}{\textwidth}
    \includegraphics[width = \textwidth]{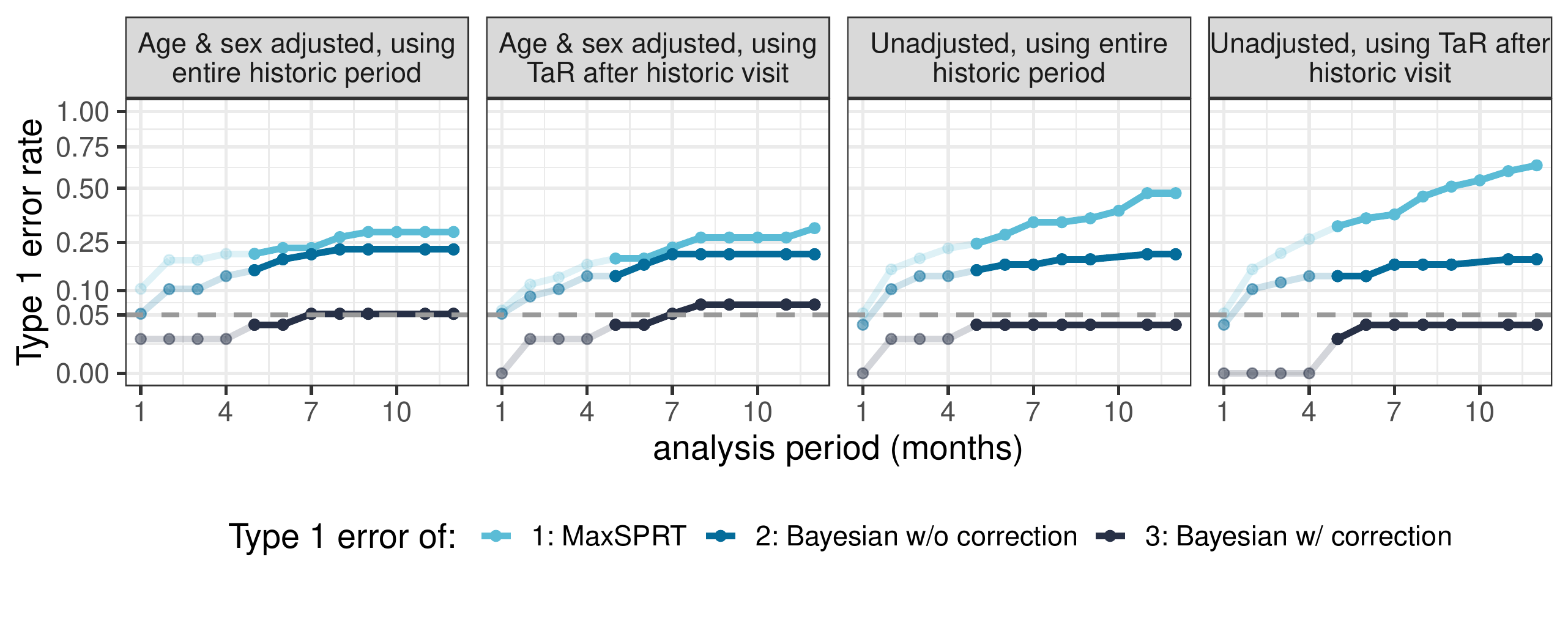}
    \vspace*{-0.4in}
    \caption{Type 1 error rates for historical comparator designs. Each panel shows results for a different design variant.}
\end{subfigure}
~
\vspace{0.1in}

\begin{subfigure}{\textwidth}
    \includegraphics[width = \textwidth]{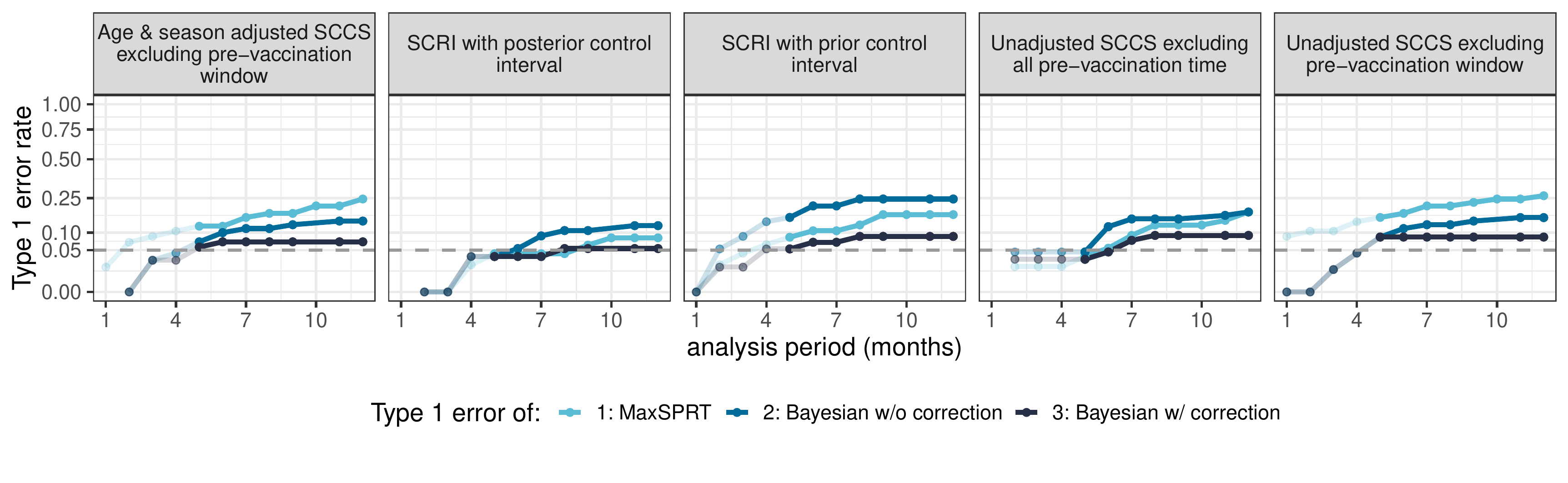}
    \vspace*{-0.3in}
    \caption{Type 1 error rates for SCCS designs. Each panel shows results for a different design variant.}
\end{subfigure}
\caption{\textbf{The Bayesian framework offers better Type 1 error control than MaxSPRT.} 
Empirical Type 1 error rates over analysis period (in months) using proposed Bayesian methods and MaxSPRT. Results are shown for HPV vaccine exposure (after 1st or 2nd dose of Gardasil 9) within the CCAE database, with 1-28 days post vaccination considered as the time-at-risk.}
\label{fig: type1-comparison-HPV}
\end{figure}

In Figure~\ref{fig: type1-comparison-HPV}, as a typical example, we plot the empirical Type 1 error rates over analysis time-points (in months) using the Bayesian framework and MaxSPRT and 
for the HPV (Gardasil 9) vaccine exposures within the CCAE database with 1-28 days post vaccination as the time-at-risk.
Within the Bayesian framework, we have also implemented a baseline with the bias correction component removed (dark blue curves); for both the baseline and the full approach with bias correction, we adopt a decision threshold $\delta_1 = 0.95$. 
We remind the reader that
\emph{a lower Type 1 error rate is desired} as one generally wants to avoid producing excessive false positive signals. 
%
Further, we have intentionally made results from the first 4 months more transparent to highlight later analysis periods, since in the early phases data remain insufficient to deliver stable estimates. 

It is clear that the proposed sequential analysis procedure with BBC (dark curves) offers substantially better control of Type 1 error rates compared to MaxSPRT (light blue curves). 
This is shared between historical comparator designs (subplot (a)) and SCCS designs (subplot (b)), but the difference is more prominent with historical comparator designs --- note that for the two unadjusted designs the Type 1 error rates that MaxSPRT achieves even exceed 50\%, 10 times higher than the pre-specified $\alpha = 5\%$. 
Even though the Bayesian framework does not rely on a pre-specified significance level, with adaptive bias correction, it actually provides a near $5\%$ actual Type 1 error rate across design choices; in many cases, Bayesian sequential analysis even without bias correction still out-performs MaxSPRT in this metric, although the benefit is not as large as those offered by the full Bayesian approach.

\subsection{The Bayesian framework does \emph{not} sacrifice power}

We next examine the statistical power in identifying true safety signals. 
%
Since our Bayesian approach and MaxSPRT return notably different Type 1 error rates and we wish to make more informative comparisons, we first numerically select a decision threshold under each epidemiological design such that the end-of-analysis Type 1 error rate that the Bayesian approach returns approximately equals to the end-of-analysis Type 1 error rate from MaxSPRT. 
In this way, we are comparing their statistical power when the same amount of Type 1 error is allowed between the Bayesian approach and MaxSPRT. 

\begin{figure}
    \centering
    \includegraphics[width = 0.85\textwidth]{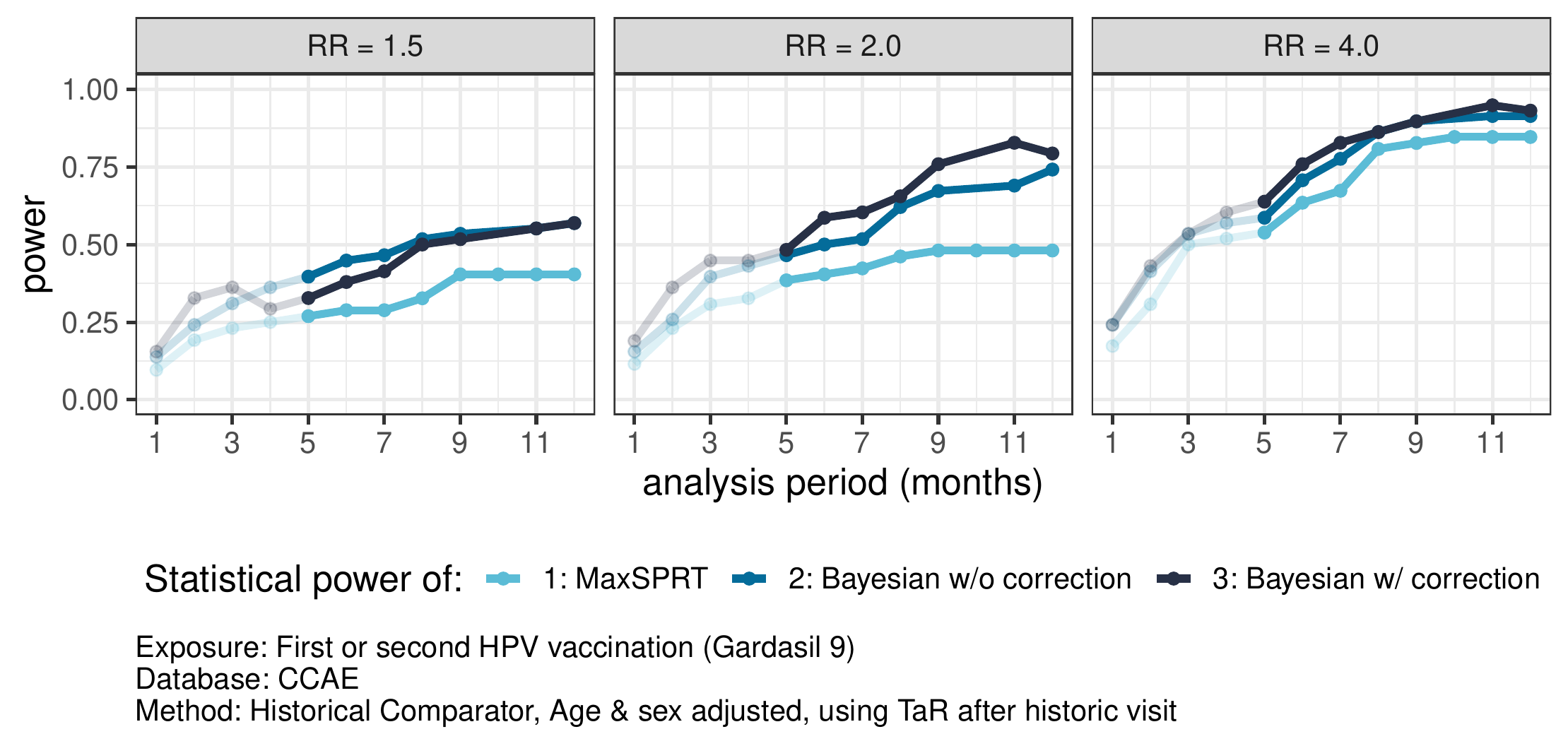}
    \includegraphics[width = 0.85\textwidth]{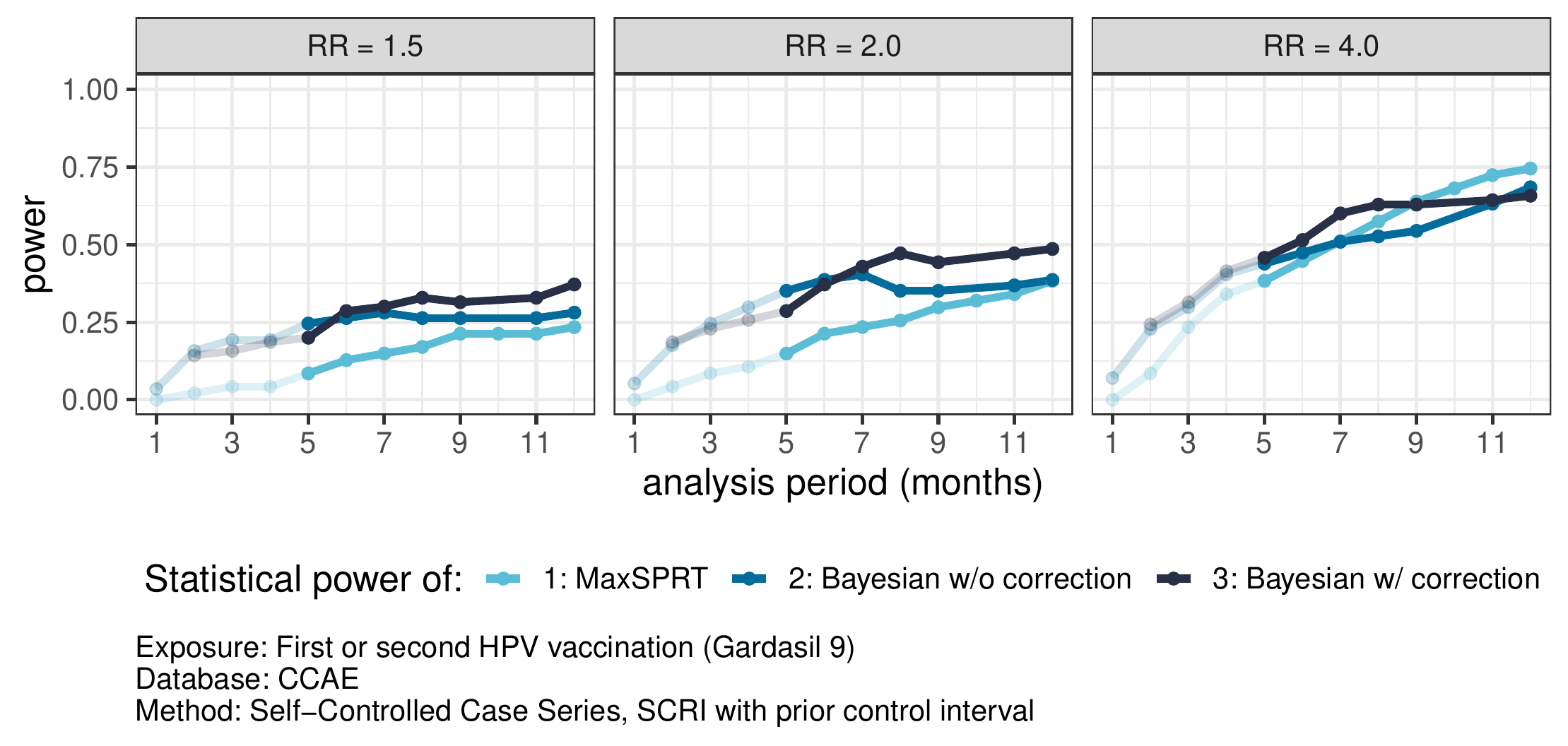}
    \caption{\textbf{Bayesian framework offers higher statistical power}. 
    Statistical power ($= 1 - $ Type 2 error rate) over analysis periods (in months) using the proposed Bayesian methods and MaxSPRT, stratified by effect sizes of positive control outcomes, with $RR = 1.5, 2$ and $4$. 
    Results are shown for HPV vaccine exposure using one representative historical comparator design (\textbf{Top}) and one representative SCCS design (\textbf{Bottom}) within the CCAE database. 
    The Bayesian method with bias correction almost consistently produces higher statistical power over MaxSPRT, when the same empirical Type 1 error level is allowed across methods.}
    \label{fig:power-comparisons}
\end{figure}

In Figure \ref{fig:power-comparisons}, we present results for one typical historical comparator design and one SCCS design in Figure~\ref{fig:power-comparisons}, with the same vaccine exposure and database as in Figure~\ref{fig: type1-comparison-HPV}. 
Despite the intrinsic trade-off between Type 1 error rate and statistical power, with the same allowance on Type 1 error, our Bayesian framework offers greater power over MaxSPRT in most scenarios, while delivering comparable power in other scenarios. 
We note that this gain is more obvious with smaller effect sizes (for example, $RR=1.5$ or $RR=2$) that, likewise, are more representative for adverse events associated with vaccination. 
This means that our proposed Bayesian approach is able to capture more true positive safety signals, while maintaining the same level of false positive decisions as MaxSPRT. 

\begin{figure}
\centering
\begin{subfigure}{0.49\textwidth}
    \includegraphics[width=\textwidth,page=2]{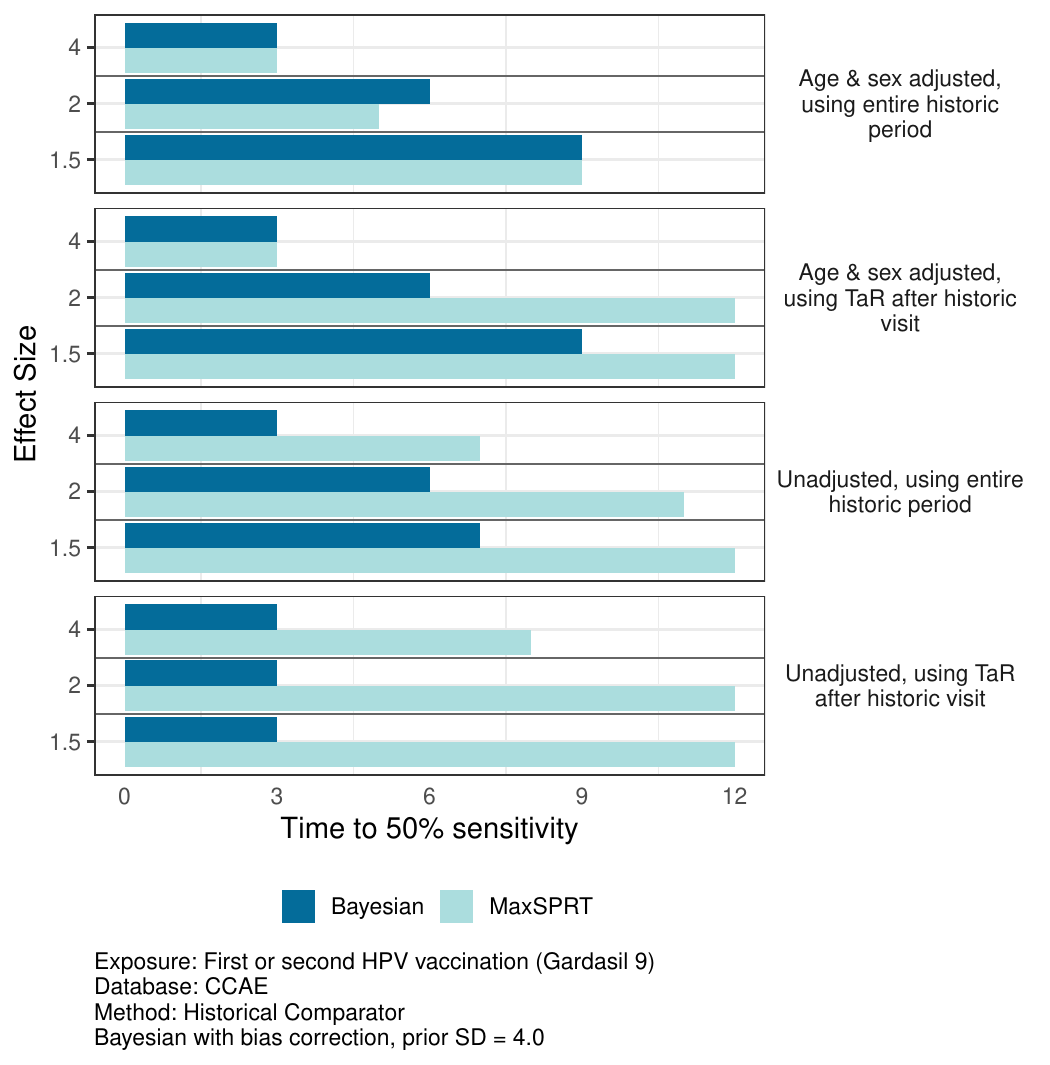}
    \caption{Historical Comparator designs.}
    \label{fig:HC-CCAE-TTS}
\end{subfigure}
\hfill
\begin{subfigure}{0.49\textwidth}
    \includegraphics[width=\textwidth,page=2]{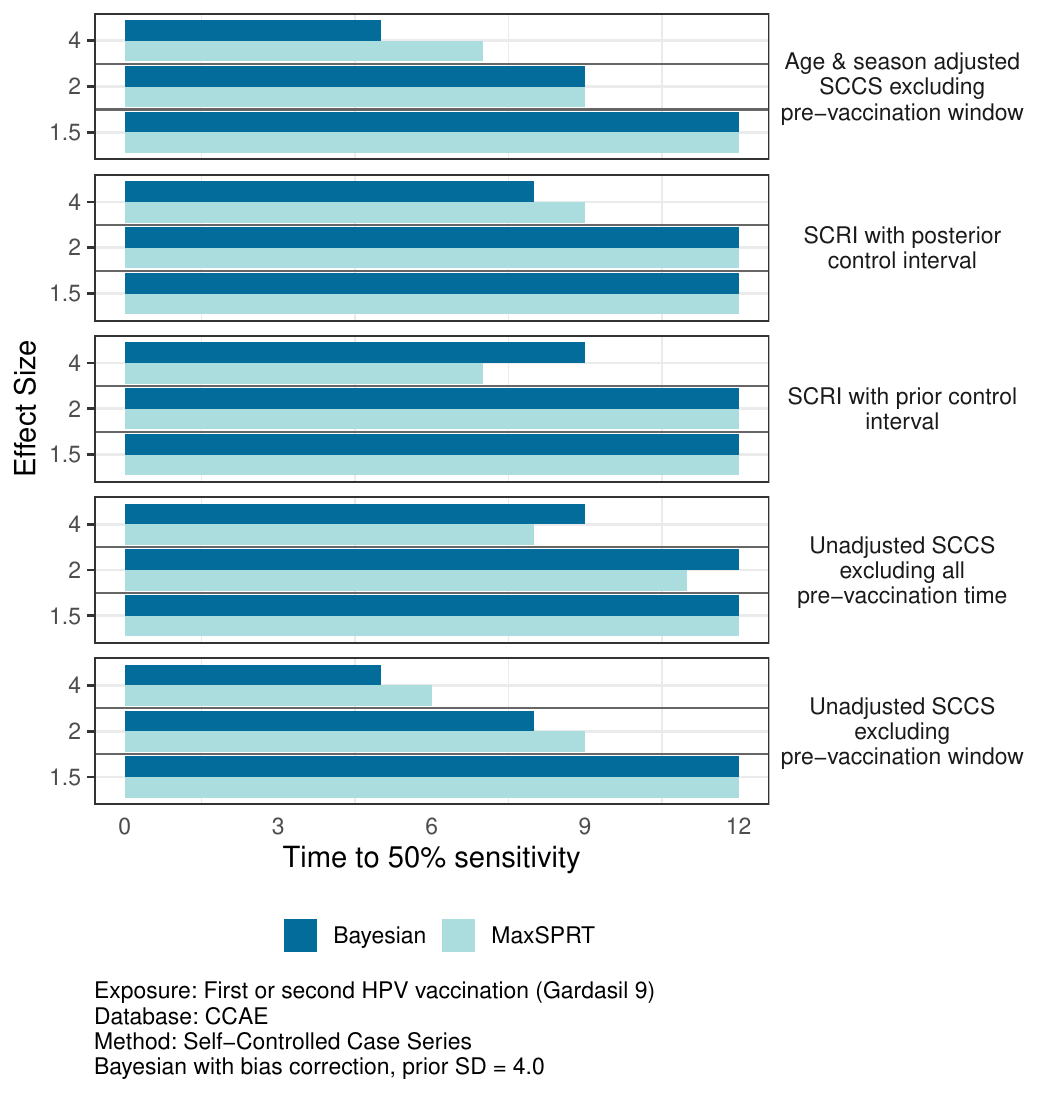}
    \caption{Self-controlled Case Series designs.}
    \label{fig:second}
\end{subfigure}
    
    \caption{\textbf{Timeliness: Bayesian framework detects true positive signals faster.} Time-to-detection it takes to reach 50\% sensitivity by the proposed BBC framework and MaxSPRT, with the same level of empirical Type 1 error rates. 
    Results are shown for HPV vaccine exposures within the CCAE database, with 1-28 days post vaccination as time-at-risk.}
    \label{fig:timeliness-comparison}
\end{figure}

We further validate this point by inspecting the timeliness of identifying true positive signals. 
In Figure~\ref{fig:timeliness-comparison}, we compare the time-to-detection, a method takes to declare safety signals for at least 50\% of all positive control outcomes, between our proposed BBC procedure and MaxSPRT, while allowing for approximately the same amount of Type 1 error.
Figure~\ref{fig: type1-comparison-HPV} presents this comparison for all epidemiological designs, again, for HPV vaccine exposure in the CCAE database with 1-28 days post vaccination as the time-at-risk.
%
Since the goal of safety surveillance is rapid detection of safety concerns, a \emph{shorter time-to-detection is desired}. 
Across almost all designs, the proposed Bayesian approach takes shorter time to detect at least 50\% of the true positive controls, particularly for smaller effect sizes. 
In few cases, the Bayesian approach does take slightly, but not substantially, longer time. 

\subsection{Bayesian bias correction yields more accurate estimation}

As the proposed Bayesian framework directly targets and corrects for estimation bias, it is able to produce more accurate and reliable estimates of log-$RR$, $\beta$. 
In Table~\ref{tab: MSEs}, we present the mean-squared errors (MSEs) in estimating $\beta$, using our proposed Bayesian BBC approach and maximum-likelihood estimation (MLE) under MaxSPRT.
For each combination of database, exposure, epidemiological design, time-at-risk choice, and true effect size $\beta$ of control outcomes, we take the average of the squared estimation errors across outcomes to produce one estimate of MSE. 
Table~\ref{tab: MSEs} summarizes the distribution of all those MSEs by examining their overall average, median, as well as the 10th, 25th, 75th and 90th percentiles.
\emph{Lower MSEs are desired} for estimation. 
It is clear that across the spectrum of all analyses, the BBC has much lower estimation error compared to MLE under MaxSPRT. 
Notably, on average, BBC yields point estimates with a nearly 80\% MSE reduction compared to MLE under MaxSPRT. 


\begin{table}
\caption{\textbf{Bayesian bias correction (BBC) provides more accurate estimation.} Summary of mean squared errors (MSEs) for estimating $\beta$, comparing BBC and MLE under MaxSPRT. 
MSEs are calculated for different effect sizes of control outcomes regarding each exposure within each epidemiological design on each database, for 3,352 total analyses where MLE under MaxSPRT is able to produce estimates. }
\label{tab: MSEs}
\centering
\begin{tabular}{p{0.06\textwidth}<{\centering}p{0.08\textwidth}<{\raggedleft}p{0.08\textwidth}<{\raggedleft}p{0.08\textwidth}<{\raggedleft}p{0.08\textwidth}<{\raggedleft}p{0.08\textwidth}<{\raggedleft}p{0.08\textwidth}<{\raggedleft\arraybackslash}}
  \toprule
  &
  \multicolumn{6}{c}{\textbf{MSE (summary quantiles)}}  \\ [0.4em]
 & \textbf{Average} & \textbf{10\%} & \textbf{25\% }& \textbf{Median} & \textbf{75\%} & \textbf{90\% }\\ 
  \midrule
BBC & 0.709 & 0.170 & 0.245 & 0.408 & 0.911 & 1.596 \\ 
  MLE & 3.394 & 0.214 & 0.340 & 0.597 & 3.218 & 8.107 \\
   \bottomrule
\end{tabular}
\end{table}

\begin{table}
\caption{\textbf{BBC produces credible intervals with substantially higher coverage rates than MLE under MaxSPRT-produced confidence intervals.} 
Summary of coverage rates of 95\% credible/confidence intervals produced by BBC and MLE under MaxSPRT in estimating $\beta$, stratified by the true effect sizes. 
Across all effect sizes, the proposed Bayesian method produces 95\% credible intervals that with close to 95\% empirical coverage rates, while 95\% confidence intervals produced by MLE under MaxSPRT tend to under-cover. 
Empirical coverage rates are calculated for different effect sizes of control outcomes regarding each exposure within each design on each database, for 3,352 total analyses where MLE under MaxSPRT is able to produce estimates.}
\label{tab: coverage-study}
\centering
\begin{tabular}{p{0.005\textwidth}<{\raggedleft}p{0.06\textwidth}<{\centering}p{0.08\textwidth}<{\raggedleft}p{0.08\textwidth}<{\raggedleft}p{0.08\textwidth}<{\raggedleft}p{0.08\textwidth}<{\raggedleft\arraybackslash}}
  \toprule
  \multicolumn{6}{c}{\textbf{Coverage rates of 95\% intervals (summary)}} \\ [0.4em]
 & & \textbf{Average} & \textbf{25\%} & \textbf{Median} & \textbf{75\%} \\ 
  \midrule  
  \multicolumn{5}{l}{\textbf{Effect size $\beta=\log(1)$}} & \\ [0.25em]
& BBC & 0.953 & 0.941 & 0.962 & 0.980 \\ 
 & MLE & 0.682 & 0.480 & 0.811 & 0.918 \\ 
   \midrule 
   \multicolumn{5}{l}{\textbf{Effect size $\beta = \log(1.5)$}} & \\ [0.25em]
  & BBC & 0.948 & 0.938 & 0.959 & 0.978 \\ 
  & MLE & 0.677 & 0.467 & 0.801 & 0.934 \\
  \midrule 
   \multicolumn{5}{l}{\textbf{Effect size $\beta=\log(1.5)$}} & \\ [0.25em]
  & BBC & 0.943 & 0.929 & 0.957 & 0.976 \\ 
  & MLE & 0.677 & 0.467 & 0.801 & 0.934 \\
  \midrule 
   \multicolumn{5}{l}{\textbf{Effect size $\beta = \log(4)$}} & \\ [0.25em]
  & BBC & 0.925 & 0.909 & 0.944 & 0.967 \\ 
  & MLE & 0.677 & 0.467 & 0.801 & 0.934 \\
   \bottomrule
\end{tabular}
\end{table}

Taking a similar approach to examine coverage rates of 95\% credible/confidence intervals, we present a summary of coverage study in Table~\ref{tab: coverage-study} in terms of the average, median, and 25th and 75th percentiles (inter-quartile range) of coverage rates calculated across all analyses. 
BBC returns 95\% credible intervals that offer close-to-95\% coverage of the true $\beta$ values very consistently. 
At the meantime, 95\% confidence intervals using MLE under MaxSPRT fail to provide nominal coverage for more than 75\% of all confidence intervals (note that all the 75th percentiles are even below 0.95). 


\section{Case study: Guillain-Barré syndrome risk post zoster vaccination}
\label{sec: case-study}

We present results from a brief case study investigating the association between occurrences of Guillain-Barré syndrome (GBS) and exposure to either of the two doses of the zoster (Shingrix) vaccine. 
Previous work \citep{goud2021risk} suggests that the zoster (Shingrix) vaccine induces an increased risk of GBS post vaccination, with an estimate of $RR = 2.84$ (95\% CI: 1.53, 5.27) using observational data from Medicare data sources. 
We wish to examine this question within the MDCR database (a subset of Medicare) by applying the same epidemiological design used in \cite{goud2021risk}, but comparing the performance of different statistical approaches, namely the proposed Bayesian framework and MaxSPRT.

Table~\ref{tab:GBS-Zoster-study} presents the final 
estimates obtained after 12 analyses based on monthly accrued data. 
Neither the Bayesian approach nor MaxSPRT has detected a positive signal that suggests zoster vaccination elevates the risk of GBS throughout the sequential analyses. 
Our Bayesian framework does return an estimate of $RR = 2.82$ that is very close to the estimate by \cite{goud2021risk}, while MLE under MaxSPRT seems to over-estimate the effect. 
However, the credible/confidence intervals are very wide, indicating inconclusive results. 
This is because the incidence counts of GBS within MDCR stand very low: only 6 total cases of GBS (defined as subjects experiencing any instance of GBS during the observation period) were present, compared to the 44 total cases in the Medicare sources used by \cite{goud2021risk}. 
Although our estimates remain inconclusive, they do highlight the potential that Bayesian safety surveillance with BBC could  replicate and validate previous findings if more data were available. 
A simple power analysis suggests that we would need at least $20$ total GBS cases in our data source to detect a statistically significant signal.

\begin{table}[H]
\caption{$RR$ estimates for occurrence of Guillain-Barré syndrome (GBS) with versus without zoster (Shingrix) vaccinations. A zoster (Shingrix) vaccine exposure is considered as taking either of the two doses. 
A SCCS design with a post-vaccination control window and covariate adjustment is applied on the MDCR (Medicare) database. 
We compare our Bayesian framework with BBC and MLE under MaxSPRT estimates against the estimates reported by \cite{goud2021risk}. 
Columns ``Risk'' and ``Control'' under ``Case Counts'' record the numbers of all GBS cases during the risk interval (1-42 days post vaccination) and during the control interval (43-183 days post vaccination). }
\label{tab:GBS-Zoster-study}
    \centering
    \begin{tabular}{p{0.25\textwidth}<{\raggedright}p{0.03\textwidth}<{\raggedleft}p{0.17\textwidth}<{\raggedright}p{0.08\textwidth}<{\raggedleft}p{0.08\textwidth}<{\raggedleft\arraybackslash}}
    \toprule
    \multirow{2}{*}{\hspace{0.02\textwidth} \textbf{Method}} & \multicolumn{2}{c}{\textbf{Estimate}} & \multicolumn{2}{c}{\textbf{Case counts}}\\[.25em]
     & \multicolumn{1}{c}{\textbf{$RR$}} & \textbf{95\% CI} & \multicolumn{1}{c}{\textbf{Risk}} & \multicolumn{1}{c}{\textbf{Control}}\\
    \midrule
    BBC (proposed) & 2.82 &(0.430, 19.0) & 2 & 4\\
    MLE  (MaxSPRT) & 4.85 & (0.562, 41.6) &  2 & 4\\
    Goud et al. (2021) & 2.84 &(1.53, 5.27) & 24 & 20\\
    \bottomrule
    \end{tabular}
    
\end{table}

\section{Discussion}
\label{sec: discussion}
This paper proposes a Bayesian sequential analysis framework for vaccine safety surveillance with adaptive bias correction. 
Our framework delivers a unified statistical solution to simultaneously provide a flexible surveillance schedule and correct for bias induced by residual systematic error in observational data. 
Our approach relies on accrued data only without the need to pre-specify a surveillance schedule, as it summarizes data evidence through posterior distributions of effect sizes and uses posterior probabilities of hypotheses for sequential testing.  
Therefore, unlike MaxSPRT, our Bayesian framework is more flexible and adaptive to practical data settings as it does not depend on \emph{a priori} predictions on data accrual behaviors.  
Furthermore, we address the challenge of residual systematic error and confounding with a joint statistical model that adaptively learns and corrects for bias by simultaneously analyzing a large set of negative control outcomes through Bayesian hierarchical modeling.
This data-driven approach enables us to substantially reduce estimation bias and remedy testing error inflation that MaxSPRT suffers from. 
Through a comprehensive empirical evaluation on six large-scale, real-world healthcare databases covering more than 360 million unique patients, 
we demonstrate that the proposed framework offers better control of Type 1 error, high statistical power, fast detection of safety signals, and more accurate and reliable estimation. 


There are, admittedly, several limitations in this work. 
First, our adaptive bias correction approach assumes exchangeable biases in the negative controls, though exchangeability is conditional on data source, epidemiological design and analysis time.
Second, we do not explicitly model time-varying risks of adverse events and do not investigate time-varying confounding, although our sequential adaptive bias correction procedure implicitly remedies this issue in part. 
Third, we have only adopted two commonly used epidemiological designs, the historical comparator and self-controlled case series, which may not be the most suitable design for vaccine safety surveillance situations with complex roll-out schedules (e.g., COVID-19 vaccines).
Finally, as misclassification of study variables is unavoidable in secondary use of healthcare data, it is possible to misclassify exposures, covariates, and outcomes; we do not expect differential misclassification, so bias will most likely be towards the null.


Nonetheless, our empirical results lead to several interesting future directions. 
First of all, with auxiliary information on correlation structures between negative control outcomes (e.g., via clinical expert review), we can extend our framework with a hierarchical mixture modeling approach to allow for non-exchangeable bias distributions. 
Secondly, since the Bayesian framework no longer requires a pre-fixed surveillance schedule, we can motivate theoretical investigation into long-term error control with infinite time horizons in sequential testing for composite hypotheses. 
Further, motivated by the inconclusive case study, we develop meta analysis approaches for synthesizing evidence across multiple data sources to increase statistical power for safety signal detection, particularly on rare safety outcomes. 
Finally, we can perform a follow-up empirical evaluation using other epidemiological designs and with more recently approved vaccines with complex roll-out schedules. 


\subsection*{Supporting Materials}
\textbf{Supporting Materials}: we include more details supplement to the main manuscript after the reference list. 

\noindent \textbf{Open-source Software}: we provide implementation of the proposed Bayesian adaptive bias correction methods in the \texttt{R} package \texttt{EvidenceSynthesis} available at \url{https://github.com/OHDSI/EvidenceSynthesis}.

\noindent \textbf{Study Protocol}: we have made details of the large-scale empirical evaluation publicly available in the study protocol at \url{https://suchard-group.github.io/Better/Protocol.html}.

\noindent \textbf{Results Explorer}: We have made all results publicly available for exploration at \url{https://data.ohdsi.org/BetterExplorer/}. 

\subsection*{Acknowledgments}

FB, AN, KK, GH and MAS receive a contract from the US Food \& Drug Administration to support this work.
Further, FB and MAS receive grants and contracts from the US National Institutes of Health (NIH), the US Department of Veterans Affairs and Janssen Research \& Development outside the scope of this work.
AN is partially supported by NIH R01 AG068002 and GH is partially supported by NIH R01 LM006910.
PBR and MJS are employees of Janssen Research \& Development and shareholders in Johnson \& Johnson. 

\bibliographystyle{chicago}
\bibliography{ref}



\begin{center}
	{\Huge 
	Supporting Materials
}
\end{center}

\renewcommand\thesection{S\arabic{section}}
\renewcommand\thefigure{S\arabic{figure}}  
\renewcommand\thetable{S\arabic{table}}  

\setcounter{section}{0}
\setcounter{figure}{0}
\setcounter{table}{0}

\section{Technical details on Bayesian bias correction}
Following Section~3.1 of the main text, we here provide details on obtaining the posterior predictive distribution $q_t(b \mid \bX_t)$ for bias term $b$. 

In our empirical evaluation we have explored a normal model and a $t$-model for the empirical bias distribution. 

Under the normal model, we assume that $b_i$, the bias for the $i$th negative control $i$, follows
\begin{align*}
	b_i &\sim N(\bar{b}, \tau^2);\\
	\bar{b} &\sim N(\mu_b, \sigma_b^2);\\
	\tau &\sim N^+(0, \sigma_\tau^2).
\end{align*}
Here $\mu_b$ and $\sigma_b^2$ are the prior mean and variance for $\bar{b}$, the ``average'' bias; $N^+(0, \sigma_\tau^2)$ represents a half-Normal distribution (i.e., Normal truncated on $(0, +\infty)$) centered at $0$ and with variance $\sigma_\tau^2$. 

Under the $t$-model, we assume the following distributions:
\begin{align*}
	b_i &\sim t_k(\bar{b}, \tau^2);\\
	\bar{b} &\sim N(\mu_b, \sigma_b^2);\\
	\tau &\sim N^+(0, \sigma_\tau^2).
\end{align*}
Here $\bar{b}$ and $\tau$ instead represent the location and scale of a $t$ distribution with $k$ degrees of freedom. 
Using a $t$ model would allow for a heavy-tailed distribution for the bias terms and better accommodate outliers and extreme values. 

In our computation we set $\mu_b = 0, \sigma_b^2 = 2$ and $\sigma_\tau^2 = 0.5$ to avoid over-estimating the scale and variability of the bias. 
For the $t$ model, we have experimented on different degrees of freedom (note that a $t$ distribution converges to normal as $k \rightarrow +\infty$) and found that $k=4$ offers good balance between allowing for heavy tails and encouraging shrinkage. 

We perform posterior inference using Markov chain Monte Carlo (MCMC) sampling, with a full implementation in the open-source \texttt{R} package \texttt{EvidenceSynthesis}. 
We run MCMC for $110,000$ total iterations with $10,000$ burn-in steps and random initializations and obtain one sample from every $100$ iterations.

\section{Supplementary information on data characteristics}

\begin{longtable}{lrrll}
	\caption{\textbf{Data characteristics by exposures and databases.} Column ``Exposure Subjects'' shows the total number of unique people with vaccination exposure considered in the analysis. 
		Column ``Exposure Days'' shows the cumulative at-risk days for all exposure subjects. 
		Column ``Outcome counts'' shows the median event counts across 93 negative control outcomes during all exposure time periods; numbers in the parentheses are the 25th and 75th percentiles. 
		Column ``Incidence rates'' shows the incidence rate (incident count divided by exposure time) across all 93 negative control outcomes, similarly with the median and IQR. 
		Summary is presented for design choices with 1-42 days after vaccine exposure considered as the ``time-at-risk''. }\\
	\label{tab:data-characteristics}
	
	\small 
	\multirow{2}{*}{\textbf{Database}} & \textbf{Exposure} & \textbf{Exposure} & \textbf{Outcome counts} & \textbf{Incidence rates} ($\times 10^{-4}$) \\
	& \textbf{Subjects} & \textbf{Person-years} & \textbf{(median \& IQR)} & \textbf{(median \& IQR)} \\
	\midrule \multicolumn{5}{l}{\textbf{H1N1 vaccination}} \\ [0.25em] CCAE &   753,592 &  83,278.37 & 15.0 [8.0, 63.0] & 1.80 [0.96, 7.56] \\ 
	CUIMC &    10,611 &   1,155.18 & 4.0 [1.5, 10.5] & 34.63 [12.99, 90.90] \\ 
	MDCD &   206,865 &  22,770.72 & 5.0 [3.0, 18.8] & 2.20 [1.32, 8.23] \\ 
	MDCR &    12,913 &   1,453.52 & 3.0 [2.0, 5.0] & 20.64 [13.76, 34.40] \\ 
	Clinformatics &   457,565 &  50,969.72 & 18.5 [7.0, 50.8] & 3.63 [1.37, 9.96] \\ 
	Optum EHR &   156,467 &  17,913.74 & 6.0 [3.0, 18.0] & 3.35 [1.67, 10.05] \\ 
	\midrule \multicolumn{5}{l}{\textbf{Seasonal flu vaccination (Fluvirin)}} \\ [0.25em] CCAE &   119,186 &  13,420.61 & 5.0 [2.0, 14.0] & 3.73 [1.49, 10.43] \\ 
	CUIMC &       230 &      25.85 & 1.0 [1.0, 1.5] & 386.92 [386.92, 580.38] \\ 
	MDCD &    15,282 &   1,729.63 & 3.0 [2.0, 9.0] & 17.34 [11.56, 52.03] \\ 
	MDCR &       822 &      91.39 & 1.0 [1.0, 2.0] & 109.42 [109.42, 218.84] \\ 
	Clinformatics &   189,184 &  21,452.01 & 11.0 [4.0, 21.0] & 5.13 [1.86, 9.79] \\ 
	Optum EHR &    14,706 &   1,677.26 & 2.0 [1.0, 4.0] & 11.92 [5.96, 23.85] \\ 
	\midrule \multicolumn{5}{l}{\textbf{Seasonal flu vaccination (Fluzone)}} \\ [0.25em] CCAE &       957 &     101.29 & 1.0 [1.0, 1.0] & 98.73 [98.73, 98.73] \\ 
	CUIMC &     3,397 &     390.33 & 1.0 [1.0, 4.0] & 25.62 [25.62, 102.48] \\ 
	MDCD &     3,357 &     383.91 & 2.0 [1.0, 3.0] & 52.10 [26.05, 78.14] \\ 
	MDCR &    34,414 &   3,897.18 & 3.0 [2.0, 7.0] & 7.70 [5.13, 17.96] \\ 
	Clinformatics &   798,816 &  91,546.16 & 55.0 [18.5, 151.0] & 6.01 [2.02, 16.49] \\ 
	Optum EHR &   337,218 &  34,865.18 & 17.0 [6.0, 38.0] & 4.88 [1.72, 10.90] \\ 
	\midrule \multicolumn{5}{l}{\textbf{Seasonal flu vaccination (All)}} \\ [0.25em] CCAE & 3,516,811 & 397,381.22 & 115.5 [36.8, 302.5] & 2.91 [0.92, 7.61] \\ 
	CUIMC &   117,339 &  13,359.51 & 14.0 [5.0, 42.0] & 10.48 [3.74, 31.44] \\ 
	MDCD & 1,237,934 & 140,144.88 & 61.5 [24.8, 202.5] & 4.39 [1.77, 14.45] \\ 
	MDCR &   264,636 &  29,834.67 & 24.5 [9.0, 63.8] & 8.21 [3.02, 21.37] \\ 
	Clinformatics & 3,399,471 & 386,968.78 & 246.5 [86.8, 510.0] & 6.37 [2.24, 13.18] \\ 
	Optum EHR & 2,536,334 & 283,676.07 & 147.0 [53.0, 388.0] & 5.18 [1.87, 13.68] \\ 
	\midrule \multicolumn{5}{l}{\textbf{First HPV vaccination (Gardasil 9)}} \\ [0.25em] CCAE &   376,341 &  41,719.07 & 7.0 [2.0, 26.0] & 1.68 [0.48, 6.23] \\ 
	CUIMC &     7,670 &     847.18 & 1.0 [1.0, 1.5] & 11.80 [11.80, 17.71] \\ 
	MDCD &   236,683 &  26,255.02 & 4.0 [1.0, 15.8] & 1.52 [0.38, 6.00] \\ 
	Clinformatics &   173,228 &  19,114.12 & 7.0 [1.0, 11.8] & 3.66 [0.52, 6.15] \\ 
	Optum EHR &   233,985 &  25,515.60 & 6.0 [2.0, 18.5] & 2.35 [0.78, 7.25] \\ 
	MDCR & 0 & 0 & N.A. & N.A. \\ 
	\midrule \multicolumn{5}{l}{\textbf{Second HPV vaccination (Gardasil 9)}} \\ [0.25em] CCAE &    49,283 &   5,030.75 & 2.0 [1.0, 6.0] & 3.98 [1.99, 11.93] \\ 
	CUIMC &     1,172 &     121.46 & 1.0 [1.0, 1.2] & 82.33 [82.33, 102.92] \\ 
	MDCD &    15,065 &   1,527.26 & 2.5 [1.0, 4.0] & 16.37 [6.55, 26.19] \\ 
	Clinformatics &    21,377 &   2,170.16 & 1.0 [1.0, 3.0] & 4.61 [4.61, 13.82] \\ 
	Optum EHR &    28,336 &   2,894.69 & 2.5 [1.8, 4.0] & 8.64 [6.05, 13.82] \\ 
	MDCR & 0 & 0 & N.A. & N.A. \\ 
	\midrule \multicolumn{5}{l}{\textbf{First or second HPV vaccination (Gardasil 9)}} \\ [0.25em] CCAE &   378,052 &  46,749.82 & 6.5 [2.0, 28.0] & 1.39 [0.43, 5.99] \\ 
	CUIMC &     7,726 &     968.66 & 1.0 [1.0, 3.0] & 10.32 [10.32, 30.97] \\ 
	MDCD &   237,455 &  27,782.29 & 4.0 [1.0, 17.0] & 1.44 [0.36, 6.12] \\ 
	Clinformatics &   174,692 &  21,284.29 & 6.5 [1.8, 13.0] & 3.05 [0.82, 6.11] \\ 
	Optum EHR &   234,518 &  28,410.31 & 5.0 [1.5, 20.0] & 1.76 [0.53, 7.04] \\ 
	MDCR & 0 & 0 & N.A. & N.A. \\ 
	\midrule \multicolumn{5}{l}{\textbf{First Zoster vaccination (Shingrix)}} \\ [0.25em] CCAE &   148,190 &  16,204.57 & 6.0 [3.0, 19.0] & 3.70 [1.85, 11.73] \\ 
	CUIMC &    11,182 &   1,228.74 & 3.0 [1.0, 6.5] & 24.42 [8.14, 52.90] \\ 
	MDCD &    11,407 &   1,258.84 & 2.0 [1.0, 7.0] & 15.89 [7.94, 55.61] \\ 
	MDCR &    52,789 &   5,839.67 & 4.0 [2.0, 13.0] & 6.85 [3.42, 22.26] \\ 
	Clinformatics &   229,463 &  25,248.18 & 14.0 [5.0, 35.0] & 5.54 [1.98, 13.86] \\ 
	Optum EHR &   219,665 &  23,818.92 & 11.0 [3.5, 36.0] & 4.62 [1.47, 15.11] \\ 
	\midrule \multicolumn{5}{l}{\textbf{Second Zoster vaccination (Shingrix)}} \\ [0.25em] CCAE &    72,063 &   7,330.99 & 3.0 [2.0, 7.0] & 4.09 [2.73, 9.55] \\ 
	CUIMC &     4,229 &     445.03 & 2.0 [1.0, 3.2] & 44.94 [22.47, 73.03] \\ 
	MDCD &     5,379 &     562.30 & 1.0 [1.0, 4.0] & 17.78 [17.78, 71.14] \\ 
	MDCR &    30,218 &   3,109.77 & 3.0 [2.0, 9.5] & 9.65 [6.43, 30.55] \\ 
	Clinformatics &   119,556 &  12,196.95 & 8.0 [2.8, 20.2] & 6.56 [2.25, 16.60] \\ 
	Optum EHR &    63,464 &   6,609.61 & 5.0 [2.0, 12.0] & 7.56 [3.03, 18.16] \\ 
	\midrule \multicolumn{5}{l}{\textbf{First or second Zoster vaccination (Shingrix)}} \\ [0.25em] CCAE &   149,219 &  23,535.55 & 9.0 [4.0, 22.0] & 3.82 [1.70, 9.35] \\ 
	CUIMC &    11,211 &   1,673.80 & 3.0 [1.0, 5.8] & 17.92 [5.97, 34.35] \\ 
	MDCD &    11,556 &   1,821.14 & 3.0 [2.0, 9.5] & 16.47 [10.98, 52.17] \\ 
	MDCR &    53,384 &   8,949.44 & 6.0 [2.0, 19.0] & 6.70 [2.23, 21.23] \\ 
	Clinformatics &   232,669 &  37,445.11 & 21.0 [8.0, 56.0] & 5.61 [2.14, 14.96] \\ 
	Optum EHR &   220,106 &  30,428.53 & 13.0 [4.5, 45.0] & 4.27 [1.48, 14.79] \\ 
	\bottomrule
\end{longtable}

\begin{figure}[H]
	\centering
	\includegraphics[width = 0.96\textwidth]{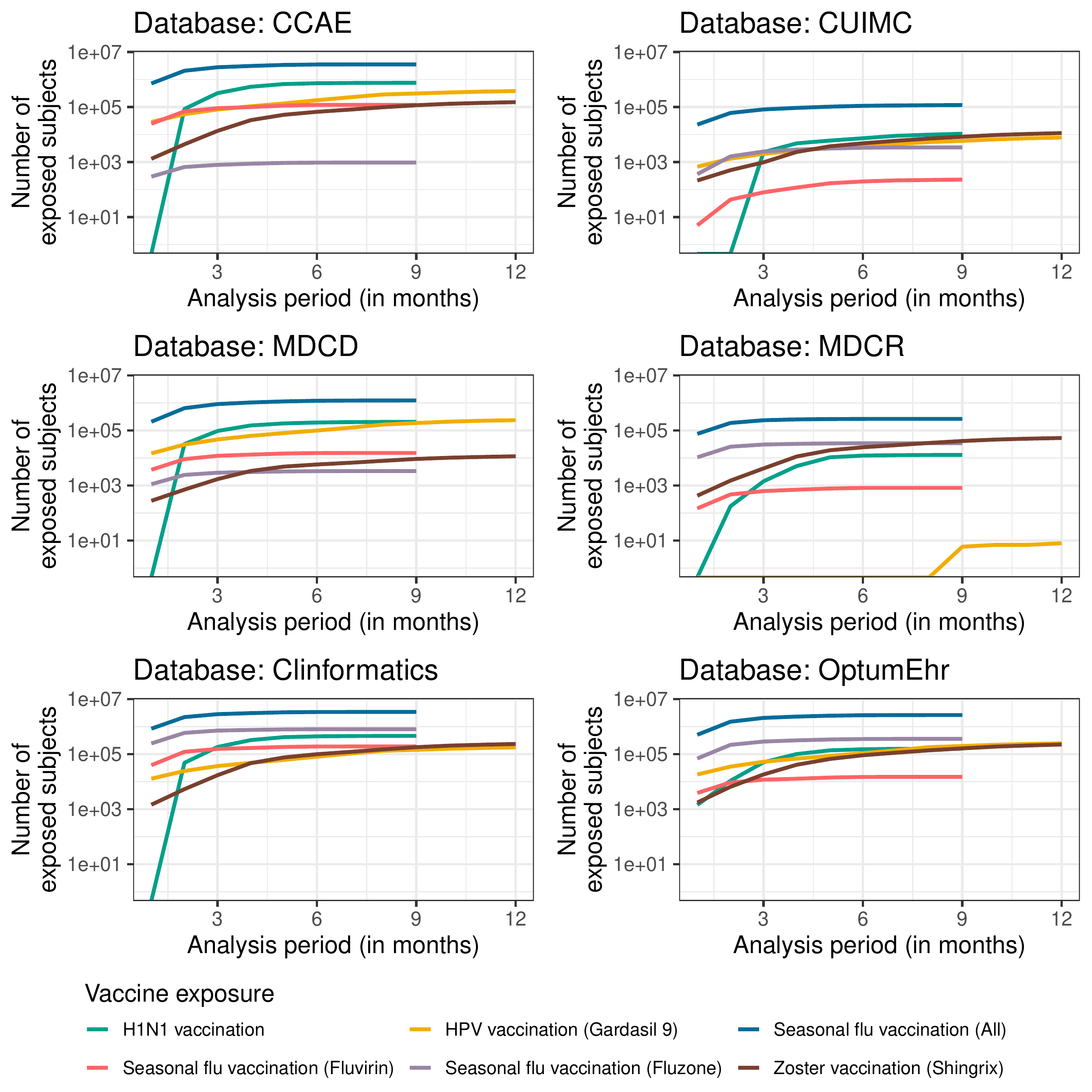}
	\caption{\textbf{Data accrual in sequential analysis periods across exposures and databases.} 
		Total number of subjects with exposure to each vaccine in each database over analysis periods (in months). }
	\label{fig:total-exposure-subjects}
\end{figure}

\end{document}


\maketitle

\renewcommand\thesection{S\arabic{section}}
\renewcommand\thefigure{S\arabic{figure}}  
\renewcommand\thetable{S\arabic{table}}

\section{Technical details on Bayesian bias correction}
Following Section~3.1 of the main text, we here provide details on obtaining the posterior predictive distribution $q_t(b \mid \bX_t)$ for bias term $b$. 

In our empirical evaluation we have explored a normal model and a $t$-model for the empirical bias distribution. 

Under the normal model, we assume that $b_i$, the bias for the $i$th negative control $i$, follows
\begin{align*}
    b_i &\sim N(\bar{b}, \tau^2);\\
    \bar{b} &\sim N(\mu_b, \sigma_b^2);\\
    \tau &\sim N^+(0, \sigma_\tau^2).
\end{align*}
Here $\mu_b$ and $\sigma_b^2$ are the prior mean and variance for $\bar{b}$, the ``average'' bias; $N^+(0, \sigma_\tau^2)$ represents a half-Normal distribution (i.e., Normal truncated on $(0, +\infty)$) centered at $0$ and with variance $\sigma_\tau^2$. 

Under the $t$-model, we assume the following distributions:
\begin{align*}
    b_i &\sim t_k(\bar{b}, \tau^2);\\
    \bar{b} &\sim N(\mu_b, \sigma_b^2);\\
    \tau &\sim N^+(0, \sigma_\tau^2).
\end{align*}
Here $\bar{b}$ and $\tau$ instead represent the location and scale of a $t$ distribution with $k$ degrees of freedom. 
Using a $t$ model would allow for a heavy-tailed distribution for the bias terms and better accommodate outliers and extreme values. 

In our computation we set $\mu_b = 0, \sigma_b^2 = 2$ and $\sigma_\tau^2 = 0.5$ to avoid over-estimating the scale and variability of the bias. 
For the $t$ model, we have experimented on different degrees of freedom (note that a $t$ distribution converges to normal as $k \rightarrow +\infty$) and found that $k=4$ offers good balance between allowing for heavy tails and encouraging shrinkage. 

We perform posterior inference using Markov chain Monte Carlo (MCMC) sampling, with a full implementation in the open-source \texttt{R} package \texttt{EvidenceSynthesis}. 
We run MCMC for $110,000$ total iterations with $10,000$ burn-in steps and random initializations and obtain one sample from every $100$ iterations.

\section{Supplementary information on data characteristics}

    \begin{longtable}{lrrll}
    \caption{\textbf{Data characteristics by exposures and databases.} Column ``Exposure Subjects'' shows the total number of unique people with vaccination exposure considered in the analysis. 
    Column ``Exposure Days'' shows the cumulative at-risk days for all exposure subjects. 
    Column ``Outcome counts'' shows the median event counts across 93 negative control outcomes during all exposure time periods; numbers in the parentheses are the 25th and 75th percentiles. 
    Column ``Incidence rates'' shows the incidence rate (incident count divided by exposure time) across all 93 negative control outcomes, similarly with the median and IQR. 
    Summary is presented for design choices with 1-42 days after vaccine exposure considered as the ``time-at-risk''. }\\
    \label{tab:data-characteristics}
    
    \small 
   \multirow{2}{*}{\textbf{Database}} & \textbf{Exposure} & \textbf{Exposure} & \textbf{Outcome counts} & \textbf{Incidence rates} ($\times 10^{-4}$) \\
    & \textbf{Subjects} & \textbf{Person-years} & \textbf{(median \& IQR)} & \textbf{(median \& IQR)} \\
       \midrule \multicolumn{5}{l}{\textbf{H1N1 vaccination}} \\ [0.25em] CCAE &   753,592 &  83,278.37 & 15.0 [8.0, 63.0] & 1.80 [0.96, 7.56] \\ 
  CUIMC &    10,611 &   1,155.18 & 4.0 [1.5, 10.5] & 34.63 [12.99, 90.90] \\ 
  MDCD &   206,865 &  22,770.72 & 5.0 [3.0, 18.8] & 2.20 [1.32, 8.23] \\ 
  MDCR &    12,913 &   1,453.52 & 3.0 [2.0, 5.0] & 20.64 [13.76, 34.40] \\ 
  Clinformatics &   457,565 &  50,969.72 & 18.5 [7.0, 50.8] & 3.63 [1.37, 9.96] \\ 
  Optum EHR &   156,467 &  17,913.74 & 6.0 [3.0, 18.0] & 3.35 [1.67, 10.05] \\ 
  \midrule \multicolumn{5}{l}{\textbf{Seasonal flu vaccination (Fluvirin)}} \\ [0.25em] CCAE &   119,186 &  13,420.61 & 5.0 [2.0, 14.0] & 3.73 [1.49, 10.43] \\ 
  CUIMC &       230 &      25.85 & 1.0 [1.0, 1.5] & 386.92 [386.92, 580.38] \\ 
  MDCD &    15,282 &   1,729.63 & 3.0 [2.0, 9.0] & 17.34 [11.56, 52.03] \\ 
  MDCR &       822 &      91.39 & 1.0 [1.0, 2.0] & 109.42 [109.42, 218.84] \\ 
  Clinformatics &   189,184 &  21,452.01 & 11.0 [4.0, 21.0] & 5.13 [1.86, 9.79] \\ 
  Optum EHR &    14,706 &   1,677.26 & 2.0 [1.0, 4.0] & 11.92 [5.96, 23.85] \\ 
  \midrule \multicolumn{5}{l}{\textbf{Seasonal flu vaccination (Fluzone)}} \\ [0.25em] CCAE &       957 &     101.29 & 1.0 [1.0, 1.0] & 98.73 [98.73, 98.73] \\ 
  CUIMC &     3,397 &     390.33 & 1.0 [1.0, 4.0] & 25.62 [25.62, 102.48] \\ 
  MDCD &     3,357 &     383.91 & 2.0 [1.0, 3.0] & 52.10 [26.05, 78.14] \\ 
  MDCR &    34,414 &   3,897.18 & 3.0 [2.0, 7.0] & 7.70 [5.13, 17.96] \\ 
  Clinformatics &   798,816 &  91,546.16 & 55.0 [18.5, 151.0] & 6.01 [2.02, 16.49] \\ 
  Optum EHR &   337,218 &  34,865.18 & 17.0 [6.0, 38.0] & 4.88 [1.72, 10.90] \\ 
  \midrule \multicolumn{5}{l}{\textbf{Seasonal flu vaccination (All)}} \\ [0.25em] CCAE & 3,516,811 & 397,381.22 & 115.5 [36.8, 302.5] & 2.91 [0.92, 7.61] \\ 
  CUIMC &   117,339 &  13,359.51 & 14.0 [5.0, 42.0] & 10.48 [3.74, 31.44] \\ 
  MDCD & 1,237,934 & 140,144.88 & 61.5 [24.8, 202.5] & 4.39 [1.77, 14.45] \\ 
  MDCR &   264,636 &  29,834.67 & 24.5 [9.0, 63.8] & 8.21 [3.02, 21.37] \\ 
  Clinformatics & 3,399,471 & 386,968.78 & 246.5 [86.8, 510.0] & 6.37 [2.24, 13.18] \\ 
  Optum EHR & 2,536,334 & 283,676.07 & 147.0 [53.0, 388.0] & 5.18 [1.87, 13.68] \\ 
  \midrule \multicolumn{5}{l}{\textbf{First HPV vaccination (Gardasil 9)}} \\ [0.25em] CCAE &   376,341 &  41,719.07 & 7.0 [2.0, 26.0] & 1.68 [0.48, 6.23] \\ 
  CUIMC &     7,670 &     847.18 & 1.0 [1.0, 1.5] & 11.80 [11.80, 17.71] \\ 
  MDCD &   236,683 &  26,255.02 & 4.0 [1.0, 15.8] & 1.52 [0.38, 6.00] \\ 
  Clinformatics &   173,228 &  19,114.12 & 7.0 [1.0, 11.8] & 3.66 [0.52, 6.15] \\ 
  Optum EHR &   233,985 &  25,515.60 & 6.0 [2.0, 18.5] & 2.35 [0.78, 7.25] \\ 
  MDCR & 0 & 0 & N.A. & N.A. \\ 
  \midrule \multicolumn{5}{l}{\textbf{Second HPV vaccination (Gardasil 9)}} \\ [0.25em] CCAE &    49,283 &   5,030.75 & 2.0 [1.0, 6.0] & 3.98 [1.99, 11.93] \\ 
  CUIMC &     1,172 &     121.46 & 1.0 [1.0, 1.2] & 82.33 [82.33, 102.92] \\ 
  MDCD &    15,065 &   1,527.26 & 2.5 [1.0, 4.0] & 16.37 [6.55, 26.19] \\ 
  Clinformatics &    21,377 &   2,170.16 & 1.0 [1.0, 3.0] & 4.61 [4.61, 13.82] \\ 
  Optum EHR &    28,336 &   2,894.69 & 2.5 [1.8, 4.0] & 8.64 [6.05, 13.82] \\ 
  MDCR & 0 & 0 & N.A. & N.A. \\ 
  \midrule \multicolumn{5}{l}{\textbf{First or second HPV vaccination (Gardasil 9)}} \\ [0.25em] CCAE &   378,052 &  46,749.82 & 6.5 [2.0, 28.0] & 1.39 [0.43, 5.99] \\ 
  CUIMC &     7,726 &     968.66 & 1.0 [1.0, 3.0] & 10.32 [10.32, 30.97] \\ 
  MDCD &   237,455 &  27,782.29 & 4.0 [1.0, 17.0] & 1.44 [0.36, 6.12] \\ 
  Clinformatics &   174,692 &  21,284.29 & 6.5 [1.8, 13.0] & 3.05 [0.82, 6.11] \\ 
  Optum EHR &   234,518 &  28,410.31 & 5.0 [1.5, 20.0] & 1.76 [0.53, 7.04] \\ 
  MDCR & 0 & 0 & N.A. & N.A. \\ 
  \midrule \multicolumn{5}{l}{\textbf{First Zoster vaccination (Shingrix)}} \\ [0.25em] CCAE &   148,190 &  16,204.57 & 6.0 [3.0, 19.0] & 3.70 [1.85, 11.73] \\ 
  CUIMC &    11,182 &   1,228.74 & 3.0 [1.0, 6.5] & 24.42 [8.14, 52.90] \\ 
  MDCD &    11,407 &   1,258.84 & 2.0 [1.0, 7.0] & 15.89 [7.94, 55.61] \\ 
  MDCR &    52,789 &   5,839.67 & 4.0 [2.0, 13.0] & 6.85 [3.42, 22.26] \\ 
  Clinformatics &   229,463 &  25,248.18 & 14.0 [5.0, 35.0] & 5.54 [1.98, 13.86] \\ 
  Optum EHR &   219,665 &  23,818.92 & 11.0 [3.5, 36.0] & 4.62 [1.47, 15.11] \\ 
  \midrule \multicolumn{5}{l}{\textbf{Second Zoster vaccination (Shingrix)}} \\ [0.25em] CCAE &    72,063 &   7,330.99 & 3.0 [2.0, 7.0] & 4.09 [2.73, 9.55] \\ 
  CUIMC &     4,229 &     445.03 & 2.0 [1.0, 3.2] & 44.94 [22.47, 73.03] \\ 
  MDCD &     5,379 &     562.30 & 1.0 [1.0, 4.0] & 17.78 [17.78, 71.14] \\ 
  MDCR &    30,218 &   3,109.77 & 3.0 [2.0, 9.5] & 9.65 [6.43, 30.55] \\ 
  Clinformatics &   119,556 &  12,196.95 & 8.0 [2.8, 20.2] & 6.56 [2.25, 16.60] \\ 
  Optum EHR &    63,464 &   6,609.61 & 5.0 [2.0, 12.0] & 7.56 [3.03, 18.16] \\ 
  \midrule \multicolumn{5}{l}{\textbf{First or second Zoster vaccination (Shingrix)}} \\ [0.25em] CCAE &   149,219 &  23,535.55 & 9.0 [4.0, 22.0] & 3.82 [1.70, 9.35] \\ 
  CUIMC &    11,211 &   1,673.80 & 3.0 [1.0, 5.8] & 17.92 [5.97, 34.35] \\ 
  MDCD &    11,556 &   1,821.14 & 3.0 [2.0, 9.5] & 16.47 [10.98, 52.17] \\ 
  MDCR &    53,384 &   8,949.44 & 6.0 [2.0, 19.0] & 6.70 [2.23, 21.23] \\ 
  Clinformatics &   232,669 &  37,445.11 & 21.0 [8.0, 56.0] & 5.61 [2.14, 14.96] \\ 
  Optum EHR &   220,106 &  30,428.53 & 13.0 [4.5, 45.0] & 4.27 [1.48, 14.79] \\ 
   \bottomrule
   \end{longtable}

\begin{figure}[H]
    \centering
    \includegraphics[width = 0.96\textwidth]{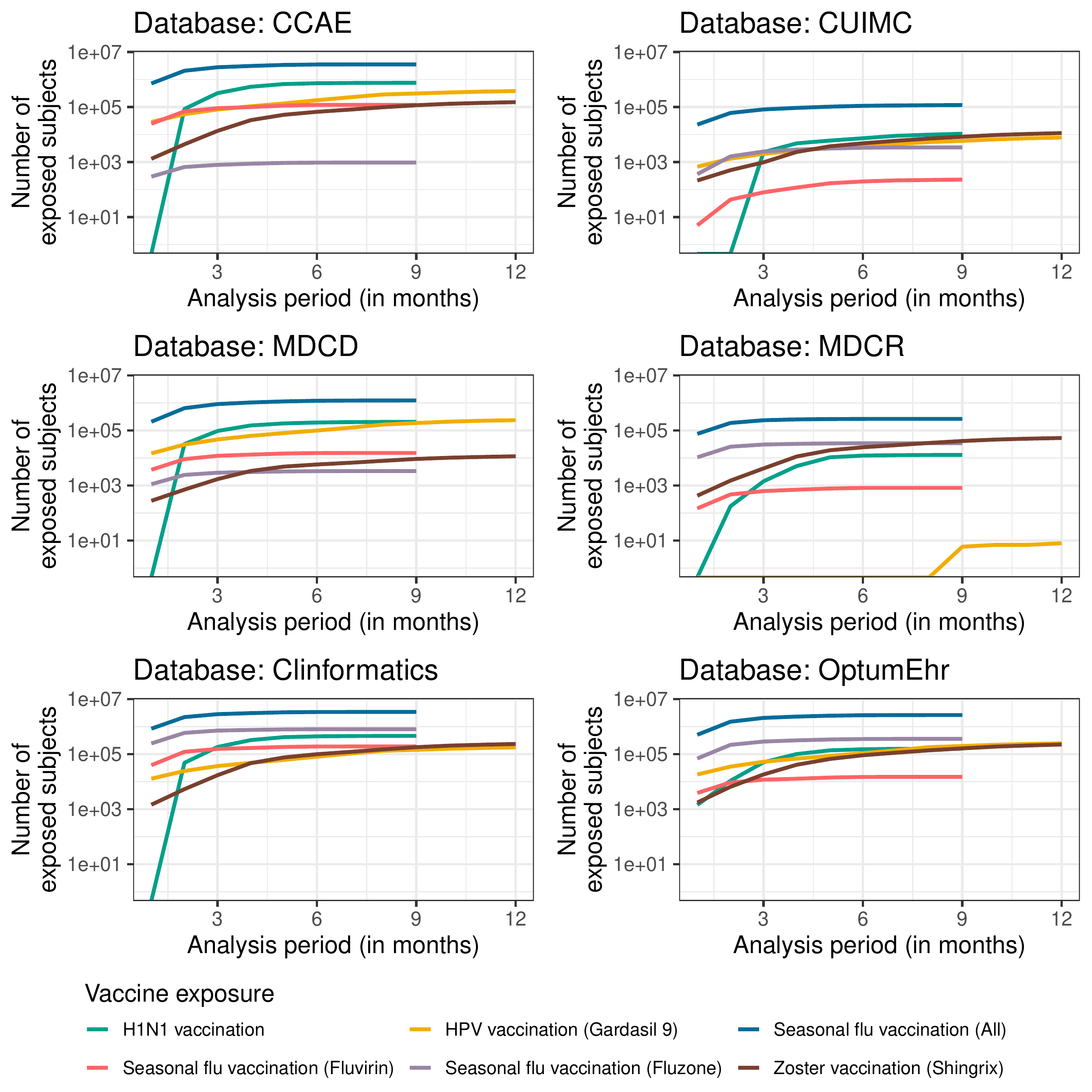}
    \caption{\textbf{Data accrual in sequential analysis periods across exposures and databases.} 
    Total number of subjects with exposure to each vaccine in each database over analysis periods (in months). }
    \label{fig:total-exposure-subjects}
\end{figure}
